%% file: main_arxiv.tex
\documentclass[sigplan,10pt]{acmart}
\usepackage{pgfplots}
\usepackage{algorithm}
\usepackage{algpseudocode}
\usepackage{amsmath}
\usepackage{multirow}
\usepackage{makecell}
\usepackage{enumerate}
\usepackage{framed}
\usepackage{textcomp}  % For \texttimes (cross mark)
\usepackage{booktabs}
\usepackage{svg}
\usepackage{subcaption}
\usepackage{soul}
\newcommand{\xmark}{\texttimes}  % Define cross mark

\renewcommand\footnotetextcopyrightpermission[1]{}

\usepgfplotslibrary{groupplots}
\usetikzlibrary{pgfplots.statistics}
\usetikzlibrary{patterns}
\pgfplotsset{width=10cm,compat=1.9}

\AtBeginDocument{%
  }

\begin{document}

\newcommand{\system}{Zorse\ }
\newcommand{\systemns}{Zorse}
\newcommand{\rt}[1]{\textcolor{red}{#1}}
\newcommand{\bt}[1]{\textcolor{blue}{#1}}
\newcommand{\gt}[1]{\textcolor{olive}{#1}}
\newcommand{\kd}[1]{\textcolor{violet}{#1}}

%%
%% The "title" command has an optional parameter,
%% allowing the author to define a "short title" to be used in page headers.
\title{\systemns: Optimizing LLM Training Efficiency on\\ Heterogeneous GPU Clusters}
%%
%% The "author" command and its associated commands are used to define
%% the authors and their affiliations.
%% Of note is the shared affiliation of the first two authors, and the
%% "authornote" and "authornotemark" commands
%% used to denote shared contribution to the research.

\author{Runsheng Benson Guo}
\affiliation{%
  \institution{University of Waterloo}
  \city{Waterloo}
  \country{Canada}
}
\email{r9guo@uwaterloo.ca}

\author{Utkarsh Anand}
\affiliation{%
  \institution{University of Waterloo}
  \city{Waterloo}
  \country{Canada}
}
\email{utkarsh.anand@uwaterloo.ca}

\author{Khuzaima Daudjee}
\affiliation{%
  \institution{University of Waterloo}
  \city{Waterloo}
  \country{Canada}
}
\email{khuzaima.daudjee@uwaterloo.ca}

\author{Rathijit Sen}
\affiliation{%
  \institution{Microsoft}
  \city{Redmond}
  \country{USA}
}
\email{rathijit.sen@microsoft.com}

%%
%% By default, the full list of authors will be used in the page
%% headers. Often, this list is too long, and will overlap
%% other information printed in the page headers. This command allows
%% the author to define a more concise list
%% of authors' names for this purpose.
% \renewcommand{\shortauthors}{Trovato et al.}

%%
%% The abstract is a short summary of the work to be presented in the
%% article.
\begin{abstract}
  Large language models (LLMs) require vast amounts of GPU compute to train, but limited availability and high 
  costs of GPUs make homogeneous clusters impractical for many organizations. 
  Instead, assembling heterogeneous clusters by pooling together GPUs of different generations
  allow them to achieve higher aggregate compute and make use of all available GPUs.
  However, training on heterogeneous clusters presents several challenges, including load balancing across GPUs, 
  optimizing memory usage to accommodate varying memory capacities, and ensuring communication-efficient 
  training over diverse network interconnects potentially spanning multiple datacenters.
  In this paper, we make the case that efficient training on heterogeneous clusters requires 
  (1) the integration of pipeline parallelism and data parallelism in a manner that is both communication- and memory-efficient, and
  (2) a more adaptable configuration of pipeline and data parallelism, which includes the capability to 
  flexibly partition GPUs into asymmetric pipeline parallel stages and to
  incorporate heterogeneous GPUs within the same data parallelism group.
  We propose \systemns, the first system to unify all these capabilities while 
  incorporating a planner that automatically configures training strategies 
  for a given workload. Our evaluation shows that \system significantly outperforms 
  state-of-the-art systems in heterogeneous training scenarios.
\end{abstract}
% \received{20 February 2007}
% \received[revised]{12 March 2009}
% \received[accepted]{5 June 2009}

%%
%% This command processes the author and affiliation and title
%% information and builds the first part of the formatted document.
\settopmatter{printfolios=true}
\settopmatter{printacmref=false}
\maketitle
\pagestyle{plain}

% \vspace{-9pt}
\vspace{-3pt}
\section{Introduction}
Large Language Models (LLMs) are built on the transformer architecture and self-attention mechanism \cite{vaswani2017attention},
which effectively captures contextual dependencies in language data. These models achieve state-of-the-art 
performance in NLP tasks, including language generation, translation, and summarization, 
making them highly versatile and widely adopted.
LLMs are trained using variants of stochastic gradient descent (SGD) \cite{sgd_ref}, which involves multiple iterations of forward passes, backward passes, and 
optimizer updates. In the forward pass, a sampled batch of inputs is fed into the model to generate an output. The backward pass 
then computes the gradients of the model's parameters with respect to the loss, a measure of the discrepancy between the model's 
outputs and the expected values. In the final step, the optimizer updates the model's parameters using the computed gradients.

The training process involves substantial matrix multiplications, which, although computationally intensive, are highly parallelizable. 
As a result, training is often accelerated using GPUs, leveraging their thousands of cores for parallel processing. 
Given the high memory and computational demands of training LLMs, the process is typically distributed across a cluster of GPUs.
% \subsection{Distributed Training Strategies}
Distributed training strategies can be broadly categorized into data parallelism and model parallelism. 
In data parallelism, each GPU processes a separate batch of data using the full model. Techniques like ZeRO-3 \cite{zero3,fsdp}
reduce memory overhead by sharding training states (parameters, gradients, optimizer state)
rather than replicating them across GPUs.

Model parallelism partitions the model itself across multiple GPUs. 
Pipeline model parallelism \cite{huang2019gpipe, pipedream} partitions a model into stages of consecutive model layers
that are assigned to different GPUs. The global batch of inputs is divided into smaller microbatches 
that are pipelined through the stages to parallelize computation. 
Rather than dividing the layers of the model, tensor model parallelism \cite{shazeer2018mesh,shoeybi2019megatron}
divides the computation within each layer across GPUs. This splits the computation at a finer granularity but 
requires extra communication to aggregate results between layers.

Some systems combine both data and model parallelism to enhance training efficiency \cite{megatron-lm, zheng2022alpa, pipedream, megascale}. 
For example, Megatron-LM \cite{megatron-lm} employs heuristics to select the most appropriate parallelization strategies: 
tensor parallelism (TP) is used within nodes with fast interconnects, pipeline parallelism (PP) is applied across nodes with slower connections, 
and data parallelism (DP) is used for further scaling.

% \subsection{Training on Heterogeneous GPU Clusters}
Distributed training systems typically assume deployment on clusters of homogeneous GPUs, which simplifies the parallelism configuration. 
Since each GPU has the same compute and memory capacity, they can be assigned an equal share of the workload. However, many machine learning practitioners 
do not have access to large, homogeneous clusters due to frequent GPU release cycles, limited cloud availability, 
and GPU shortages \cite{hap, cephalo, demystifying-het-inference, hexiscale, metis, ding2021hetseq, cannikin}. High-end GPUs are expensive, and most organizations cannot afford to 
purchase a new cluster every year to keep up with the rapid release of new GPUs. Instead, they often incrementally acquire GPUs, 
resulting in clusters of heterogeneous GPUs over time. Additionally, several studies have highlighted the 
limited availability of GPUs in the cloud \cite{cephalo, demystifying-het-inference, parcae,cross-region}. Newer GPU models are rarely accessible, and it is often difficult to reserve more than 32 GPUs, even for older 
generations \cite{cephalo, demystifying-het-inference}. Nevertheless, users can still reserve larger GPU clusters for computation by combining GPUs 
of different generations. 

For these reasons, clusters of heterogeneous GPUs are becoming more and more ubiquitous, and there has been lots of interest in 
developing efficient distributed training techniques for heterogeneous clusters \cite{hap, cephalo, hexiscale, metis, park2020hetpipe}.
However, training on such clusters introduces several challenges: 

\noindent\textbf{(1) Diverse GPU capabilities}: Different GPUs vary in compute power and memory capacity. Hence, efficient training 
requires flexible parallelism strategies tailored to each device.

\noindent\textbf{(2) Network heterogeneity}: Variability in interconnect bandwidth (e.g., inter- vs. intra-node, inter- vs. intra-datacenter) requires communication-efficient 
parallelism techniques that support large differences in link speeds.

\noindent\textbf{(3) Memory constraints}: GPUs with lower memory capacity demand memory-efficient training strategies to maximize compute utilization without running out of memory.

To address these challenges, we introduce \systemns, the first system to resolve all of these issues simultaneously. 
\system combines PP and DP by partitioning the model into stages, 
pipelining computation across them, and applying DP within each stage. \system employs ZeRO-2 DP, 
which reduces memory consumption by sharding optimizer states and gradients. Unlike existing PP approaches, 
each stage is divided into ministages, computations are interleaved to avoid fully materializing stage parameters and gradients in memory, 
Moreover, both activations and parameters are offloaded to CPU memory when not in use, significantly reducing memory overhead.
\system also supports heterogeneous PP, allowing different stages to use varying numbers of GPUs, and 
each stage to be parallelized with different GPU configurations.
This paper makes the following contributions:
\begin{enumerate}
    \item We examine key challenges in heterogeneous LLM training and provide insights on designing efficient systems for this use case.
    \item We built \systemns, a flexible system that simultaneously addresses these challenges through a novel and efficient integration 
    of pipeline and data parallelism.
    \item We design a planner that efficiently navigates the large search space of configurations in \systemns, automatically optimizing 
    training for a given workload and cluster.
    \item We demonstrate that \system greatly outperforms existing heterogeneous training systems, 
    achieving up to 3x higher training throughput on three representative clusters for models scaling up to 65 billion parameters.
    % \item We make \system available as open-source software\footnote{We will add the link to the repo in the final version of the paper.} 
    % for adoption and future extension by the community.
\end{enumerate}

\section{Challenges in Heterogeneous Training}
\label{sec:challenges}
In this section, we provide insights and establish key challenges to overcome for achieving efficient training on heterogeneous GPU clusters.

\input{plots/tp_vs_fsdp}

\subsection{Tensor Parallelism in Heterogeneous Clusters}
\label{subsec:tp_het}
Tensor parallelism (TP) is typically deployed only in clusters of high-end GPU machines interconnected by 
very high bandwidth technologies such as NVSwitch \cite{megatron-lm}. 
This is because TP requires frequent all-to-all communication between model layers. 
Without extremely high-bandwidth interconnects, the overhead from these communications becomes prohibitively large, 
despite optimizations that overlap communication with computation \cite{async_tp}. 
In Figure \ref{fig:tp_vs_fsdp}, we compare the GPU utilization when training  
with TP versus data parallelism (DP) across common AWS VMs. 
The plot shows that GPU utilization with TP is low relative to DP for most VMs, and is only comparable to DP
on the 8$\times$A100 VM that has high bandwidth NVSwitch interconnects.
Moreover, TP is generally advantageous only in the following training scenarios where DP or PP become impractical:

\noindent\textbf{(1) Exceptionally large models}: 
TP is useful for training very large models with layers that exceed the GPU's memory capacity, as it partitions the computation within each layer.

\noindent\textbf{(2) Extremely large batch sizes}: When DP leads to excessively large batch sizes that impact training convergence, 
TP can be used to distribute training across more GPUs without further increasing the batch size.

These scenarios, however, are not typical in heterogeneous training environments. Such environments often utilize low- or mid-tier 
GPUs with limited networking bandwidth, since high-end GPUs are limited in availability \cite{cephalo}. 
Additionally, organizations using heterogeneous clusters are typically resource constrained,
and therefore would not be training at the scale of models and batch sizes that require TP.

\begin{framed}
  \textbf{Takeaway \#1}: TP is unsuitable for heterogeneous training due to its high communication overhead
  and misalignment with heterogeneous training scenarios.
\end{framed}
\vspace{-3pt}

\input{plots/network_bandwidth}

\subsection{Data Parallelism in Heterogeneous Networks}
In homogeneous clusters with high-bandwidth interconnects, DP has been shown to 
scale efficiently to large clusters of over 1000 GPUs \cite{zero++, zero3}. 
However, the scalability of DP is limited in heterogeneous clusters due to substantial network variability caused by 
differences in (1) intra-node networking, (2) inter-node networking, and (3) intra- and inter-datacenter networking.
Figure \ref{fig:network_bandwidth} illustrates the network heterogeneity in bandwidth between common GPUs/VMs on AWS.

DP uses collective operations like AllReduce, which synchronize model updates across GPUs using all-to-all communication patterns. 
However, these collectives utilize links inefficiently in heterogeneous networks and
can significantly increase communication latencies since faster links are bottlenecked by slower links.
Moreover, inter-datacenter bandwidth is typically insufficient to train efficiently with DP \cite{cross-region}.
Instead of relying on DP alone, DP can be combined with pipeline parallelism (PP) which requires 
significantly less communication and can be used over slower inter-node or inter-datacenter links \cite{cross-region, megatron-lm}.

\begin{framed}
  \textbf{Takeaway \#2}: Efficiently scaling training over heterogeneous networks
  with slow links necessitates a combination of data parallelism and pipeline parallelism.
\end{framed}

\input{plots/zero2_vs_zero3}
\vspace{-10pt}

\subsection{PP + DP in Heterogeneous Clusters}
\label{subsec:pp_dp_het}
ZeRO-3 \cite{zero3, zero++} (or FSDP \cite{fsdp} in PyTorch) is a popular variant of DP designed to avoid 
redundant storage of training states across GPUs. While this method significantly reduces the memory requirement for 
storing training states by a factor proportional to the number of GPUs in the cluster, it also increases communication 
overhead. Specifically, ZeRO-3 must gather parameter shards before each forwards and backwards computation (parameters are 
sharded afterwards). This overhead is exacerbated when
ZeRO-3 is combined with PP, as PP divides a batch of inputs into microbatches that are pipelined to overlap computation 
across pipeline stages. Consequently, the frequency of parameter gathering increases proportionally with the number of microbatches, 
creating a significant communication bottleneck and slowing down training unless extremely fast interconnects are available, 
which is uncommon in heterogeneous clusters.

There have been proposals to amortize the communication cost of gathering parameters in ZeRO-3 
by swapping the order of computation \cite{cephalo,lamy2021layered}.
Instead of processing all layers for one microbatch before moving to the next, these approaches sequentially process 
all microbatches for one layer before proceeding to the next layer. While this strategy reduces the communication overhead 
when combining ZeRO-3 and PP, it introduces a pipelining overhead: subsequent pipeline stages cannot begin processing until 
the previous stage reaches the last layer, limiting the ability to overlap computations across stages.

Alternatively, PP can be combined with ZeRO-2, which shards optimizer state and gradients but keeps a full set 
of parameters in memory. This avoids the need to gather parameters before each computation, achieving higher training throughput
but significantly increases memory usage, as shown in Figure \ref{fig:zero2_vs_zero3}.
This is especially problematic in heterogeneous clusters, where some GPUs can be memory constrained relative to 
their compute capacity \cite{cephalo}. In Figure \ref{fig:zero2_vs_zero3}, although both V100s and T4s have the same memory, V100s are more than three times faster. 
As model size increases, V100s in PP + ZeRO-2 are unable to process a workload proportional to its compute capacity due to memory constraints, 
resulting in lower training throughput.

\begin{framed}
  \textbf{Takeaway \#3}: Existing variants of DP cannot be combined with PP without
  introducing significant communication, pipelining, or memory overhead.
\end{framed}
\vspace{-3pt}

\subsection{Symmetric Parallelism Configuration}
Most state-of-the-art distributed training systems \cite{torchtitan, rasley2020deepspeed, galvatron, megatron-lm}
integrate PP with DP by organizing GPUs into a 2D mesh, with PP along one axis
and DP on the other. This configuration simplifies system design and narrows the parallelism plan search space.
However, in heterogeneous clusters, this approach may overlook efficient training configurations. Such clusters often
contain asymmetry, with variations in GPU count per node, hardware, and datacenter location. Efficient training
configurations would ideally group GPUs of the same datacenter, node, or GPU type for DP, while applying PP across
different groups, ensuring that GPUs within each DP group share similar networking and computational capabilities.
Moreover, GPUs may need to be grouped into varying group sizes to accomodate differences in memory capacity (e.g. 
GPUs with less memory require smaller groups to reduce the memory overhead of parameter replication).
Supporting such configurations is important for heterogeneous training,
and requires a system capable of asymmetrically partitioning GPUs into PP stages of different sizes.

\begin{framed}
  \textbf{Takeaway \#4}: Supporting PP with variably-sized DP stages
  is essential for optimizing training on heterogeneous clusters.
\end{framed}

\subsection{Pipelining Overheads}
Training systems optimized for homogeneous clusters \cite{torchtitan, rasley2020deepspeed, galvatron, megatron-lm,zheng2022alpa} 
typically distribute batch sizes evenly across GPUs within a DP stage. This uniform allocation 
results in training bottlenecks when heterogeneous GPUs share a stage, as faster GPUs will be 
blocked waiting for slower GPUs within the same stage. Avoiding this issue in heterogeneous 
environments restricts GPUs within a DP group to be of the same type, which therefore 
scales the number of PP stages by the number of GPU types. However, adding 
more pipeline stages can introduce inefficiencies, such as increased idle time (bubbles) and 
additional inter-stage communication. Moreover, increasing the number of stages constrains the flexibility 
in distributing model layers across stages to achieve balanced computation, especially considering the 
relatively few layers in modern LLMs. Consequently, it is 
important for heterogeneous training systems to support balancing of computational loads 
across different GPUs within a stage. This capability allows for configurations with fewer stages
that incur lower pipelining overhead.

\begin{framed}
  \textbf{Takeaway \#5}: Balancing computational loads across heterogeneous GPUs within a DP stage
  facilitates training configurations with reduced pipelining overhead. 
\end{framed}
\vspace{-3pt}

\section{Limitations of Existing Systems}
\label{sec:limitations}
Given the challenges discussed above, we examine the limitations of 
representative systems that embody state-of-the-art methods for heterogeneous training.

\textbf{\textit{General 3D Parallel Frameworks.~}}
Systems such as TorchTitan \cite{torchtitan} and DeepSpeed \cite{rasley2020deepspeed} support 3D
parallelism with DP, PP, and TP, but are not explicitly designed for heterogeneous clusters.
In heterogeneous clusters, these frameworks need to be manually configured with 
asymmetric model partitions to balance computation time across stages in PP. 
However, they do not provide the flexibility to asymmetrically partition GPUs across pipeline 
stages or to use different batch sizes within DP groups to accommodate heterogeneous 
GPUs within a stage. Additionally, its DP can be optimized for either 
memory or communication efficiency, but not both simultaneously. For instance, 
in TorchTitan, DP can configured to use either ZeRO-2 which is communication efficient
but not memory efficient, or ZeRO-3 which is memory efficient but not communication efficient.

\textbf{\textit{Heterogeneity-Aware 3D Parallel Frameworks.~}}
Frameworks such as HexiScale \cite{hexiscale} and Metis \cite{metis} are designed for 3D parallelism 
in heterogeneous clusters. They support asymmetric PP and permit varying batch 
sizes within a DP group, enabling balanced computation across different GPUs. 
However, some parameters still require manual tuning, e.g., HexiScale 
requires specifying the batch size distribution for DP and microbatches for PP. 
Additionally, these frameworks do not support ZeRO-3 due to communication bottlenecks when 
integrated with PP. Instead, they employ ZeRO-2 and standard DP, respectively. 
When PP and DP exceed the cluster's memory capacity, TP is employed to reduce memory usage,
but this often leads to poor utilization in heterogeneous clusters (Section \ref{subsec:tp_het}).

\textbf{\textit{Heterogeneity-Aware ZeRO-3.~}}
Systems like Cephalo \cite{cephalo} and Poplar \cite{zhang2025poplar} utilize ZeRO-3 for training on heterogeneous clusters.
Instead of using PP or TP to scale training for larger batch sizes, Cephalo employs gradient
accumulation to reduce the memory requirements by splitting large batch sizes into
smaller microbatches. It also reorders the computation to avoid regathering parameters
for each microbatch. However, due to the lack of support for PP, Cephalo cannot efficiently utilize
network resources in highly heterogeneous networks, resulting in networking bottlenecks
when the batch size is insufficient to hide communication latencies \cite{cephalo}.
\input{tables/system_comparison}

\begin{figure*}[]
  \centering
  \includegraphics[width=\textwidth]{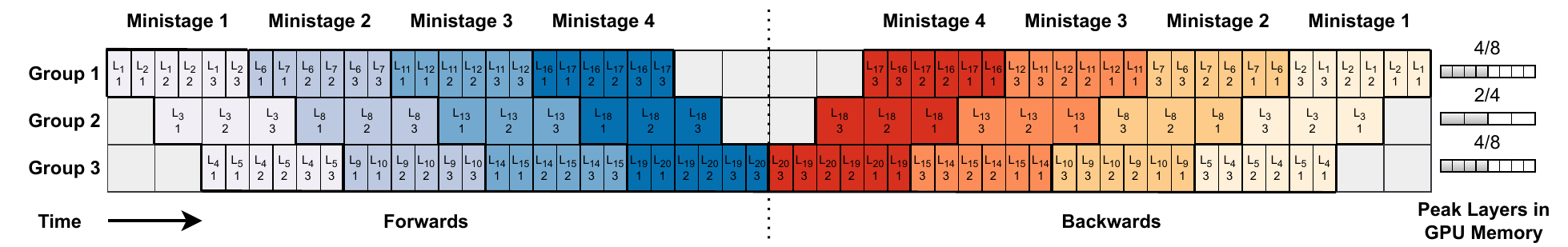}
  \caption{Interleaved pipelining in \systemns. The diagram shows a training iteration for a model with 20 layers ($L_i$) and 3 microbatches. 
  There are 3 GPU groups, each with 4 ministages. Due to different computational speeds, groups have different numbers of layers per ministage. 
  Each group maintains the current and next ministage, offloading others to CPU memory.}
  \label{fig:interleaved_pipelining}
\end{figure*}
\vspace{-23pt}

\section{Design}
To overcome the challenges in heterogeneous training discussed in Section \ref{sec:challenges} and the limitations
of existing systems in Section \ref{sec:limitations}, we developed \systemns.
Table \ref{tab:system_comparison_table} compares \system with existing systems.
\system is the first to integrate all essential capabilities for efficient training on 
heterogeneous clusters. 

\system trains LLMs with a combination of PP and ZeRO-2 DP, 
using a combination of interleaved pipelining and offloading to optimize the memory efficiency of ZeRO-2 DP
while avoiding the communication overhead of ZeRO-3 DP.
\system also supports flexibly configuring PP and DP, including asymmetric partitioning of GPUs across pipeline stages
and varying batch sizes within a DP group.
Finally, \systemns's planner automatically determines optimized training configurations 
for a given workload. In this section, we describe the components of \system in detail.

\subsection{Efficient PP and DP Integration}
Instead of conventional PP which assigns one stage per GPU, \system uses interleaved pipelining  
which assigns multiple smaller \textit{ministages} to each GPU and interleaves computation across them. However,
traditional interleaved pipelining \cite{megatron-lm} uses a 1F1B \cite{pipedream} schedule, 
interleaving computation across ministages. This means that ministage parameters cannot be
sharded without incurring extra communication overhead.

\subsubsection{Interleaved Pipelining}
\system instead uses a GPipe-style \cite{huang2019gpipe} pipelining schedule to efficiently integrate ZeRO-2 DP with PP,
 illustrated in Figure \ref{fig:interleaved_pipelining}.
The forwards pass is processed for each ministage sequentially, finishing all microbatches for a ministage before moving on to the next.
The backwards pass is processed similarly in the reverse order. Since the microbatches are all processed together for each 
ministage, we offload other ministages to CPU memory when they are not being used. 
With this design, \system needs to maintain only the parameters for 2 ministages in memory at a time
(the current ministage and the next ministage that is being prefetched in parallel), rather than all the 
parameters in the stage in standard PP + ZeRO-2. 

Let $L$ be the total number of layers in the stage, $S$ be the number of stages,
$P_{layer}$ be the number of parameters in each layer, $D_{dp}$ be the degree of DP, and $M$ be the number of microbatches. 
Table \ref{tab:memory_footprint} summarizes the parameter memory and communication requirements of \systemns, PP + ZeRO-2, and PP + ZeRO-3.
As the number of ministages increases, the memory utilization of \system approaches that of PP + ZeRO-3 and is significantly lower than PP + ZeRO-2. 
Additionally, \system has the same communication requirement as PP + ZeRO-2, with one AllGather per layer in the forwards and backwards pass. 
This is significantly lower than PP + ZeRO-3, which requires one AllGather per microbatch of each layer.

\begin{table}[h]
  \centering
  \resizebox{\columnwidth}{!}{%
  \begin{tabular}{l|c|c|c}
    \hline
    \textbf{Strategy} & \textbf{Materialized} & \textbf{Sharded} & \textbf{\# AllGathers} \\
    & \textbf{Parameters} & \textbf{Parameters} & \\
    \hline
    \system & $2 \times \frac{L}{S} \times P_{layer}$ & 0 & $2 \times L$ \\
    PP + ZeRO-2 & $L \times P_{layer}$ & 0 & $2 \times L$ \\
    PP + ZeRO-3 & $2 \times P_{layer}$ & $(L-2) \times \frac{P_{layer}}{D_{dp}}$ & $2 \times L \times M$ \\
    \hline
  \end{tabular}
  }
  \caption{Memory and communication comparison of different training strategies.}
  \label{tab:memory_footprint}
  \vspace{-15pt}
\end{table}

\subsubsection{Interleaved Optimizer Updates}
\begin{figure}[]
  \centering
  \includegraphics[width=0.5\textwidth]{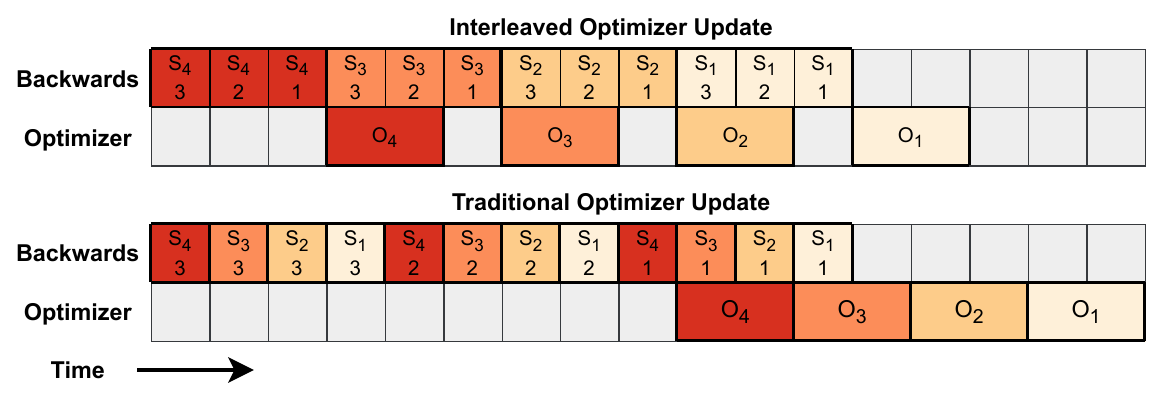}
  \caption{Interleaved optimizer updates in \system vs. traditional pipeline parallelism (4 ministages, 3 microbatches). 
  Interleaved optimizer updates free gradient memory earlier and better overlap with computation.}
  \label{fig:interleaved_optimizer_updates}
  \vspace{-8pt}
\end{figure}

\system further takes advantage of its interleaved pipelining schedule to interleave optimizer updates.
Instead of starting the optimizer update after the backwards pass completes for all layers, \system starts the optimizer update
for each ministage as soon as its own backwards pass completes, while the computation for the next ministage is simultaneously starting. 
This design (Figure \ref{fig:interleaved_optimizer_updates}) offers two key advantages:

\noindent\textbf{(1) Reduced peak memory utilization:} By freeing the gradients of each ministage as soon as its optimizer 
update is finished, \system avoids the need to retain gradients until the entire backward pass is complete.

\noindent\textbf{(2) Decreased communication overhead:} The optimizer update requires averaging gradients across GPUs 
within its DP group. By interleaving optimizer updates, this communication can overlap with the 
computation of subsequent ministages. This overlap is not possible if all optimizer updates start at the end of the pipeline schedule.

\subsubsection{Activation Checkpoining \& Offloading}
As in prior work \cite{megatron-lm,pytorch_trillion_model, hexiscale}, \system checkpoints activations after each transformer layer,
discarding activations between layer boundaries and recomputing them when they are needed in the backwards pass. 
However, the remaining activations at layer boundaries still require a significant  chunk of GPU memory due to the pipeline schedule.
Sequential processing of all forwards passes before backwards passes means that layer boundary activations for all microbatches of all layers
need to be maintained in memory. With a batch size of $B$, $L$ layers, sequences of length $S$, a hidden layer size of $H$, and $D$ bytes per parameter, 
the memory overhead is 
$B \times L $ activations of size $ S \times H \times D $ bytes. Even at smaller training scales, this can amount to many GBs of memory.

Thus, in addition to offloading parameters, \system also offloads layer boundary activations to CPU memory during the forwards pass, which are loaded back during the backwards pass. 
With offloading, \system needs to maintain activations for only the current microbatch being computed and for the next microbatch being prefetched. 

In Section \ref{sec:comp_offload_overlap}, we describe how \system efficiently overlaps offloading (and loading)
of activations, and model parameters with computation to hide offloading overhead.

\begin{figure}[]
  \centering
  \includegraphics[width=0.5\textwidth]{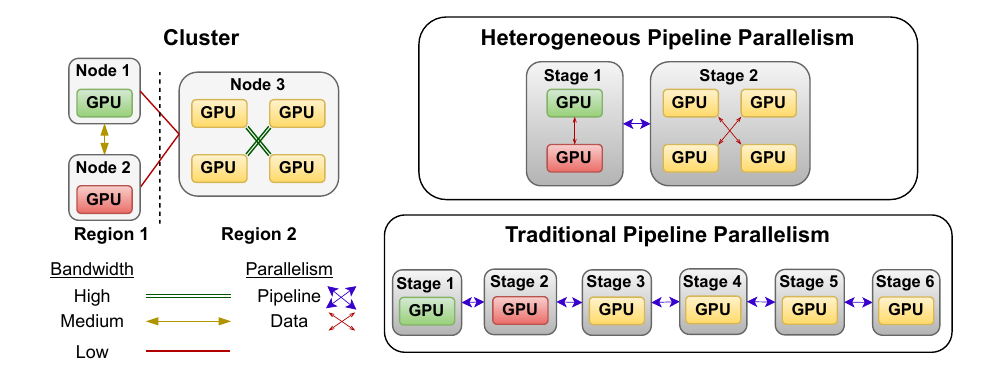}
  \caption{Heterogeneous pipeline parallelism in \system compared to traditional pipeline parallelism.}
  \label{fig:het_pp}
  \vspace{-12pt}
\end{figure}

\subsection{Heterogeneous Pipeline Parallelism}
\system supports \textit{heterogeneous pipeline parallelism}, which allows for each pipeline stage
to have: (1) a different number of GPUs, and (2) a combination of different GPU types.

This differs from traditional PP, which fixes a uniform number of GPUs per stage and the same GPU type within each stage.
This flexibility allows \system to more efficiently configure PP with DP in heterogeneous clusters, aligning
with the varying numbers of each GPU type, varying numbers of GPUs per VM, and varying memory and compute capabilities
in heterogeneous clusters.

Figure \ref{fig:het_pp} illustrates heterogeneous PP on a 3-node cluster with 1, 2, and 4 GPUs of different types across two regions. 
With heterogeneous PP, nodes in Region 1 form a single stage with 3 GPUs, while Region 2's node forms another stage with 4 GPUs. 
This approach leverages high-bandwidth intra-region connections for DP communication while using PP across the lower-bandwidth inter-region link, 
reducing pipeline stages to just two. Traditional PP, however, requires six single-GPU stages to maintain GPU uniformity, significantly increasing pipeline overhead.
Moreover, with more stages, it may become infeasible to partition the model layers in a way that evenly distributes computation across stages.
This is especially true for LLM models, which have a limited number of layers.

\textbf{\textit{Cross-stage Communication Algorithm.~}} In traditional PP, 
communication is straightforward due to the uniformity of GPUs both within and across stages, allowing for one-to-one 
communication between GPUs in different stages. However, in heterogeneous PP where the number and type of GPUs per stage varies, 
data must be reshuffled across stages, necessitating many-to-many communication patterns. 
Additionally, data must be balanced across heterogeneous GPUs within a stage to evenly distribute computation. 

\system computes an optimized communication plan to redistribute microbatches across stages and balance computation within stages. 
Microbatches are assigned to GPUs based on their relative layer runtimes. This algorithm estimates completion times for 
microbatches of the current stage and remaining compute time for each GPU in the next stage. It then assigns the $i$th 
completed microbatch to the GPU with the $i$th highest remaining runtime, prioritizing GPUs with more work to minimize overall 
stage runtime. 

\subsection{Planner}
\label{sec:planner}
\begin{figure*}[]
  \centering
  \includegraphics[width=\textwidth]{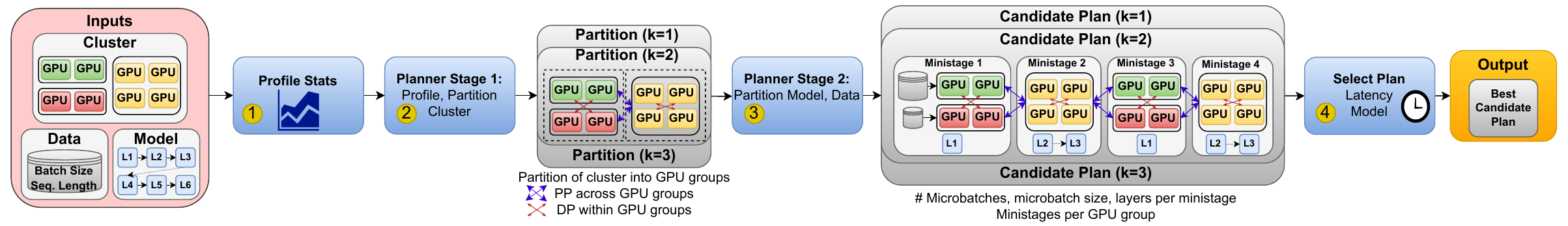}
  \caption{Architecture of \systemns's planner: \textcircled{1} profile workload and cluster, \textcircled{2} partition cluster into GPU groups, 
  \textcircled{3} partition model and data across GPU groups, and \textcircled{4} select the best plan considered.}
  \label{fig:opt_arch}
\end{figure*}
\system initially profiles the cluster and workload to gather model runtime and networking statistics. 
This data is then utilized by the planner to determine the final training configuration. 
\systemns's planner employs a two-phase optimization process to optimize the training configuration:
\begin{itemize}
  \item In Phase 1, the cluster is partitioned into \textit{GPU groups}. PP will be applied across GPU groups and 
  DP within each GPU group. The planner considers the best way to partition the cluster into $k$ GPU groups,
  for $k \in [1, N]$ where $N$ is the number of GPUs in the cluster.
  \item In Phase 2, the planner computes an optimized configuration for each cluster partition from Phase 1.
  This includes the number of microbatches, the size of each microbatch, the number of ministages per 
  GPU group, and the partitioning of the model into these ministages.
\end{itemize}
The architecture of \systemns's planner is shown in Figure \ref{fig:opt_arch}. 

\subsubsection{Profiling.}
We measure inter-node and intra-node bandwidths between pairs of nodes and GPUs, respectively, for use in cluster partitioning. 
We profile model layer runtimes on each GPU for small batch sizes and fit a linear model to predict runtimes unseen batch sizes. 
These bandwidth measurements and runtime predictions are used to estimate training latencies in Phase 2. 
To accelerate profiling, we: (1) parallelize intra-node profiling, and (2) profile inter-node bandwidths 
only between unique VM configurations, as large clusters often contain redundant configurations. 

\subsubsection{Phase 1: Cluster Partitioning.}
In the first phase, the planner determines how to divide the cluster nodes into $k$ GPU groups. GPUs within the 
same group utilize DP, while PP is applied across groups. Therefore, the goal is to create a partition that minimizes 
inter-group bandwidth usage. This results in an efficient setup that uses communication-efficient PP over lower bandwidth 
links between groups and DP over higher bandwidth links within groups.

We represent the cluster as a fully connected graph \( G = (V, E) \), where \( V \) represents the GPUs and \( E \) 
represents the edges with weights \( w(u, v) \) indicating the bandwidth between GPUs \( u \) and \( v \). 
The partitioning task is framed as a min-$k$ cut problem on this graph. Specifically, the objective is to divide 
\( V \) into \( k \) disjoint subsets \( V_1, V_2, \ldots, V_k \) such that the total weight of the edges between 
different subsets is minimized. This is called the \textit{min-k cut}:
\[
\min\sum_{\substack{u \in V_i, v \in V_j \\ i \neq j}} w(u, v),\quad \text{where} \bigcup_{i=1}^{k} V_i = V, \mkern3mu  V_i \cap V_j = \emptyset \text{ if } i \neq j
~~.\]

This partitioning minimizes the bandwidth sum between GPU groups, ensuring that higher bandwidth links are used within 
groups. This problem is solved for each $k$ from $1$ to the total number of GPUs in the cluster, $N$.

Since exact min-$k$ cut solutions have an impractically large $O(N^{k^2})$ time complexity \cite{min_kcut}, we use the SPLIT greedy approximation 
algorithm \cite{kcut_greedy}, which guarantees a solution within a factor of $2 - 2/k$ of optimal. SPLIT iteratively computes min 2-cuts of the 
current graph and removes those edges until $k$ connected components remain. Thus, we can efficiently generate approximate min $k$ cuts for all 
values from $2$ to $N$ in a single execution with $k=N$.
With each min 2-cut requiring $O(N^3)$ time \cite{StoerWagner1997} and $N$ iterations needed, the overall time complexity 
of cluster partitioning is $O(N^4)$.

\subsubsection{Phase 2: Model Configuration.}
\label{sec:phase_two}

In the second phase, the planner utilizes cluster partitions from Phase 1 to determine an optimized training configuration, 
which encompasses model layer partitioning into ministages, GPU group ordering, and microbatch size selection.
Due to the extensive search space, \system employs heuristics to prune candidate configurations before 
evaluating the remaining options with a latency model for performance assessment and a memory model to verify 
GPU memory constraints are respected.

\noindent\textbf{Heuristics.} In pipeline parallelism, the pipeline is bottlenecked by the slowest GPU group.
Thus, to minimize training latency, we need to balance runtime across GPU groups by partitioning model layers 
proportionally to each group's aggregate processing speed, estimated as the sum of layer processing rates across GPUs in the group. 
Layers assigned to each GPU group are divided into evenly sized ministages, which are then ordered across groups in round robin fashion.

The order of GPU groups impacts pipeline startup time because the first ministage must gather sharded model 
parameters without having prior ministages to overlap this communication. We order GPU groups by descending 
intra-group bandwidth to minimize initial communication delay and expedite pipeline startup. This arrangement 
enables subsequent GPU groups to better overlap their communication with computation from preceding GPU groups.

\noindent\textbf{Enumeration.} We enumerate over all possible configurations of batch size and
number of ministages per GPU group, which is at most the number of model layers. Hence, there are only
$O(B\cdot L)$ configurations to consider, where $B$ is the global batch size and $L$ the number of layers. 
We use our lightweight latency and memory model to select the configuration with the lowest estimated latency that satisfies the memory constraints.

\subsubsection{Latency Model.}
We model the total training latency as:
\begin{equation}
L_{total} = (L_{forwards} + L_{backwards}) \cdot N_{ministages} + L_{startup}
\end{equation}
Ministage forward pass latency $L_{forwards}$ encompasses 
computation time across all GPUs, AllGather communication, PP communication, and accounts for their overlap.
Ministage backward pass latency $L_{backwards}$ similarly accounts for computation time (with activation recomputation) and communication 
overhead for AllGather and ReduceScatter operations.
Pipeline startup latency $L_{startup}$ represents the time needed to initialize the pipeline, including initial parameter AllGather 
and PP communication latencies that cannot overlap with computation.

\subsubsection{Memory Model.}
We model the total per-GPU memory consumption as:
\begin{equation}
M_{total} = M_{params} + M_{grads} + M_{optim} + M_{activations}
\end{equation}
where $M_{params}$ represents memory for model parameters, $M_{grads}$ gradients, $M_{optim}$ optimizer states, and $M_{activations}$ activations. 

\section{Implementation}
In this section we provide more details on \systemns's implementation and optimizations.
\system is built on top of FSDP \cite{fsdp} (ZeRO-2 and ZeRO-3 in PyTorch). 
We integrated interleaved pipeline parallelism with FSDP, handling management of ministages,
pipelining microbatches across GPU ministages, and communication between GPU groups.

\subsection{Interleaved Pipelining}
We modified FSDP's parameter management to support interleaved pipeline parallelism by addressing key timing assumptions. 
Unlike FSDP's ZeRO-2 which keeps parameters materialized between forward and backward passes, we reshard and offload ministage 
parameters to CPU post-forward pass, then reload and unshard during the backward passes. To handle FSDP's premature resharding 
after the first microbatch's backward pass, our implementation delays resharding until all microbatches complete. We limit memory 
usage by configuring the FSDP prefetcher to fetch only the next immediate ministage. 
Finally, we create per-ministage optimizers that execute independently after their corresponding backward passes complete.

\subsection{Communication Libraries}
We use NCCL \cite{NVIDIA_NCCL_2024} for DP communication within GPU groups. For PP communication, we encountered limitations 
with NCCL's blocking P2P behavior \cite{nvidia2020p2p}, which can cause deadlocks in heterogeneous PP due to cyclic dependencies. 
We implemented a hybrid approach using NCCL where possible and GLOO P2P (non-blocking) between GPU groups
where cycles in communication are possible. While GLOO cannot utilize NVLink, this was not a performance bottleneck 
as NVLink is rare in heterogeneous clusters and typically connects GPUs within the same DP group rather than across PP groups.

\subsection{Computation-Networking Overlap}
\label{sec:comp_network_overlap}
We optimize communication-computation overlap by managing parameter sharding at the layer level 
instead of ministage level. By gathering layers sequentially rather than as a single large chunk, 
computation can begin once the first layer is ready while subsequent layers are prefetched 
in parallel. This approach extends to the backward pass, where we prefetch CPU-offloaded 
parameters ahead of when they are needed, preventing computation stalls.

\subsection{Computation-Offloading Overlap}
\label{sec:comp_offload_overlap}
PyTorch's built-in CPU offloading incurs significant performance overhead by blocking GPU 
computation and executing optimizer steps on CPU when parameter offloading is enabled. Our custom 
implementation utilizes separate CUDA streams for parameter and activation offloading without blocking 
GPU computation. Similarly, we prefetch activations and parameters in advance to prevent stalling 
during backward pass. Finally, we execute optimizer steps on GPU prior to offloading to avoid 
costly optimizer updates on CPU.

\section{Performance Evaluation}
We compare \system to state-of-the-art heterogeneous training systems across three representative 
clusters with up to 128 GPUs, and LLMs with up to 65B parameters.
Training throughput and cluster utilization is evaluated in Section \ref{sec:training_throughput} and \ref{sec:cluster_scaling}.
We evaluate \systemns's training efficiency on heterogeneous clusters in Section \ref{sec:comparison_to_homogeneous_training}, and
investigate how components of \system  such as offloading and interleaved pipelining contribute to its performance in Section \ref{sec:interleaved_pipelining_analysis}
and \ref{sec:ablation_study}. Finally, we break down the optimizer runtime in Section \ref{sec:optimizer_overhead}.

\subsection{Experimental Setup}
For all systems, we train with a global batch size of 1 million tokens using FP16 mixed precision
and apply activation checkpointing to reduce memory overhead \cite{pytorch_trillion_model,hexiscale}. 
We use standard Llama \cite{touvron2023llama} models with 7 to 65 billion parameters. 

\noindent\textbf{Clusters.~}
We evaluate on three clusters representative of common heterogeneous training scenarios.
These clusters include VMs from Azure and AWS, contain up to 128 GPUs of low-, mid-, and high-end GPUs, 
and span up to 2 regions. Table \ref{tab:gpu_specs} details GPU specifications, while 
Table \ref{tab:cluster_gpu_configs} presents the cluster configurations. We use sequence lengths of 
4096, 1024, and 512 for clusters A, B, and C respectively, training with longer sequences on clusters 
with more powerful GPUs.

\input{tables/gpu_specs}

\input{tables/cluster_gpu_configs}

\noindent\textbf{Baselines.~}
We compare against representative state-of-the-art techniques for training on heterogeneous GPU clusters:
\begin{itemize}
    \item TorchTitan-Het: 3D parallelism in TorchTitan \cite{torchtitan} adapted for heterogeneous clusters by partitioning the model
    unevenly to balance compute across PP stages. We try different combinations of 
    DP (ZeRO-2), PP, and TP, reporting the best performing configuration.
    \item  HexiScale \cite{hexiscale}: Combines DP (ZeRO-2), tensor, and pipeline parallelism, leveraging
    a planner to select an optimized training configuration.
    \item Cephalo \cite{cephalo}: Distributes compute workload and training state unevenly with FSDP to utilize
    heterogeneous resources efficiently.  
\end{itemize}

\noindent\textbf{Metrics.~}
We evaluate training performance using two key metrics:
(1) TFlops: The floating point operations per second achieved during training, which measures training throughput.
(2) Hardware FLOPS Utilization (HFU \cite{palm_hfu}): The ratio of achieved TFlops to the cluster's peak theoretical TFlops, 
which quantifies GPU utilization efficiency.

\input{tables/training_throughput_results}

\vspace{-8pt}
\subsection{Training Throughput}
\label{sec:training_throughput}
In this section we compare \system to existing systems across three representative heterogeneous training scenarios. 
A summary of the results is provided in Table \ref{table:training_throughput_results}. More details on the training configurations

\noindent\textbf{A: Small-size cluster of high-end GPUs.~}
This cluster consists of 4 H100s and 16 A100s, representative of scenarios where users have limited high-end GPUs but can combine them to 
form a larger cluster. GPUs within each node are connected via NVSwitch, while inter-node bandwidth is significantly 
lower at 50 Gbps. \system achieves superior performance over all baselines, with its relative speedup increasing with model size, 
reaching up to 3$\times$ on the LLama 65B model.

TorchTitan-Het partitions the model into 5 stages, each with 4 GPUs, one grouping the 4 H100s together and the rest grouping the A100s. 
It employs a $2\times 2$ ZeRO-2 DP$\times$TP configuration, which works well for smaller models but runs out of memory on larger ones. 
Gradient accumulation is used to manage memory constraints, but this adds communication overhead for ZeRO-2 from extra parameter gathering.

HexiScale employs PP + ZeRO-2 DP, adjusting PP and DP degrees to manage memory usage. However, larger models still face high memory pressure, 
leading to suboptimal partitioning and underutilization of H100 GPUs. Despite having nearly 3$\times$ the TFlops of A100s, H100s have only 15\% more memory, 
and are unable to train with 3$\times$ the number of layers without running out of memory.

Cephalo uses ZeRO-3 across all GPUs, but the significant disparity in intra-node and inter-node bandwidths (over 35-fold) causes AllGather 
and ReduceScatter communications to bottleneck training.

In contrast, \system maintains high efficiency even with larger models by grouping GPUs within each node for DP and applying PP across 
nodes, possible with \systemns's support for heterogeneous PP. Its memory optimizations allow it to train larger models while evenly balancing computation 
across GPUs, unlike other systems that must compromise performance to avoid running out of memory.

\noindent\textbf{B: Medium-size cluster of low, middle, \& high end GPUs.~}
This cluster comprises 8 A100s, 16 A10Gs, 16 V100s, and 24 T4s, representing scenarios where users have limited access to GPUs with diverse performance capabilities 
but aim to utilize them concurrently for training. The cluster presents challenges due to its heterogeneity in computational power, memory, and 
networking. For instance, V100s and A100s benefit from faster intra-node NVLink interconnects, whereas T4s and A10Gs rely on PCIe; 
T4s and V100s share the same memory capacity, yet V100s are twice as fast; A10Gs and V100s have similar computational speeds, but A10Gs offer 1.5 times more memory.

\system partitions the cluster into four groups, each consisting of VMs with the same GPUs and hardware. This setup facilitates ZeRO-2 DP within groups sharing the same 
networking hardware, ensuring slower interconnects do not impede faster ones. For the larger 13B and 33B models, \system employs 4 and 6 ministages per GPU, 
respectively, increasing parameter offloading and reducing memory usage. This strategy is crucial for performance, as it frees memory, 
allowing for more flexible layer distribution across GPUs. \system consistently achieves a 1.5$\times$ -- 4$\times$ speedup in training 
throughput across all models and baselines.

TorchTitan-Het and HexiScale lack efficient activation offloading mechanisms, relying instead on partitioning models into many stages with PP to manage 
memory usage. However, without interleaved pipelining, they struggle to balance computation across PP stages due to the limited number of model layers. 
For example, FlashFlex divides the LLama 33B model into over 16 stages, but with only 40 layers available, the partitioning is too coarse to 
balance computation effectively. The A100, with 5$\times$ the TFlops of the T4, would need 5$\times$ more layers to achieve balance. 
Unlike \systemns, TorchTitan-Het cannot flexibly configure PP and DP, and runs out of memory when training LLama 33B.

Cephalo efficiently manages memory by fully sharding parameters across the cluster, balancing computation across GPUs. 
However, it faces bottlenecks due to collective communication in the highly heterogeneous network.

\noindent\textbf{C: Large-size cluster of low- \& middle-end GPUs.~}
This cluster comprises 128 GPUs across two AWS regions: 16 A10Gs and 48 T4s in one, and 16 V100s and 48 T4s in the other. 
It represents scenarios where users lack high-end GPUs but have access to numerous low- and mid-tier GPUs. Training is challenging due 
to limited GPU memory ($\le$ 24GB) and high communication latencies from the large GPU count and slower cross-region links. 
Consequently, training throughput and utilization are generally lower. Nevertheless, \system consistently achieves at least a 1.5$\times$ speedup over baselines.

To prevent OOM errors on larger models, \system employs varied DP sizes tailored to each VM configuration. For instance, all A10Gs form a single DP group, 
while V100s are divided into two groups of 8 GPUs each, as they need to handle a similar layer count as A10Gs but with 50\% less memory. 
This smaller DP grouping reduces memory demands for model parameter replication. T4s, assigned fewer layers due to their slower speed, 
are split into two groups of 24 within each region to avoid cross-region DP.

TorchTitan-Het, unable to asymmetrically partition GPUs, required a higher PP degree to manage memory, leading to high pipelining overhead 
and poor performance. HexiScale's use of TP to cut memory usage resulted in high communication overhead and also led to poor performance. 
Cephalo's uneven training state partitioning allowed V100s to store less state, but the cross-region network's uneven partitioning overhead caused 
communication bottlenecks.

\input{plots/cluster_scaling}

\subsection{Cluster Scaling}
\label{sec:cluster_scaling}
We validate \systemns's scalability with increasingly heterogeneous and larger clusters. For each cluster, 
we first evaluate the training performance of the largest model that can be trained on the slowest GPUs. We then incrementally 
add faster GPUs to the training group, gradually increasing the total number of GPUs and heterogeneity until the entire cluster is 
utilized. Results are presented in Figure \ref{fig:cluster_scaling}.

Adding heterogeneous GPUs into the training group across various clusters significantly improves training throughput. Furthermore, 
cluster utilization (HFU) generally remains stable or improves, as observed in Cluster B. Improvements can be attributed to reduced 
memory requirements per GPU with the addition of more GPUs, enabling more efficient training configurations that better balance computation. 
% These results demonstrate that \system can scale effectively to large and heterogeneous clusters. 
These results demonstrate that when homogeneous GPUs are limited, 
they can be pooled together to achieve higher throughput without sacrificing training efficiency with \systemns.

\input{plots/homogeneous_comparison}

\subsection{Comparison to Homogeneous Training}
\label{sec:comparison_to_homogeneous_training}
We assess the training efficiency of \system on heterogeneous versus homogeneous clusters in Figure \ref{fig:homog_hfu}. 
For each heterogeneous cluster, we compare the HFU achieved per GPU type when training on the entire cluster against training 
on only the homogeneous subset of those GPUs. This evaluation uses the largest model that can be trained on the clusters 
without running out of memory. Homogeneous training is typically expected be more efficient due to its smaller 
scale and uniform hardware, which reduces communication overhead and eliminates the need for balancing computation. 
Nevertheless, our results demonstrate that \system consistently achieves efficiency levels comparable to homogeneous training, 
reflecting its ability to balance network, compute, and memory heterogeneity. 
This highlights \systemns's capability to effectively scale training across heterogeneous clusters, 
achieving higher throughput than possible with homogeneous clusters, while maintaining efficiency.

\input{plots/interleaved_pipelining}
\subsection{Interleaved Pipelining Analysis}
\label{sec:interleaved_pipelining_analysis}
In Section \ref{subsec:pp_dp_het}, we demonstrated how PP + ZeRO-3 is memory efficient but can lead to very low throughput due to high
communication overhead, whereas PP + ZeRO-2 is memory inefficient but can achieve high throughput due to low communication overhead.
We analyze how interleaved pipelining in \system is able to achieve a good balance between low memory and high throughput, and show the 
tradeoff as we scale the interleaving factor, corresponding to the number of ministages assigned to each GPU.

In Figure \ref{fig:interleaved_pipelining_scaling}, we evaluate the effects of interleaving using two homogeneous clusters of 16 A100s, and 16 A10Gs. 
We use a homogeneous cluster to isolate the impact of interleaving from side effects that may result from compute imbalance in heterogeneous clusters.
There is a slight drop in throughput when going from 1 ministage (no interleaving) to 2 ministages per GPU, due to increases in pipelining overhead.
However, as we increase the number of ministages past 2, the additional drop in ministages is minimal, while continuing to decrease memory utilization.
In both training setups, with maximum interleaving, we are able to reduce the memory utilization by 40$\%$ while incurring only a 20$\%$ drop in throughput.
When memory is limited in heterogeneous clusters, this is an effective tradeoff since it allows for compute to be balanced more evenly across GPUs,
which can often improve throughput and make up for the additional pipelining overhead.

We also plot the performance of PP with ZeRO-2 and ZeRO-3 for comparison. \system achieves much higher throughput than PP + ZeRO-3
with comparable memory utilization, particularly on the A10Gs (due to slower links).  
When interleaving is maximized, \system actually uses less memory since it also offloads sharded parameters.
We measured offloading overhead to be minimal, contributing at most 3\% performance impact across all workloads.

\input{plots/ablation_study}

\subsection{Ablation Study}
\label{sec:ablation_study}
We conduct an ablation study to evaluate the impact of \systemns's key components on training throughput and memory utilization. Starting with a 
baseline PP + ZeRO-2 implementation in TorchTitan, we incrementally add: (1) activation offloading (O), (2) interleaved pipelining and optimizer updates (I), and 
(3) heterogeneous pipeline parallelism (H). We evaluate these components using Llama 65B on Cluster A, with results shown in Figure \ref{fig:ablation_study}.

The PP + ZeRO-2 baseline fails to train the model due to excessive activation memory overhead. Adding activation offloading (O) enables training with 
a PP degree of 5 and ZeRO-2 degree of 4, but H100 GPUs remain underutilized despite being 3$\times$ faster than A100s, as they lack sufficient 
memory to store proportionally more layers. Interleaved pipelining (I) further reduces memory utilization by 26\% due to parameter offloading and interleaved optimizer updates. 
Finally, with heterogeneous pipeline parallelism (H), we reduce pipeline stages to 4 by grouping GPUs within each node. This configuration achieves balanced 
computation across GPUs, trading slightly higher memory utilization for significant performance improvements.

\input{plots/optimizer_time_breakdown}

\subsection{Planner Optimization Time}
\label{sec:optimizer_overhead}
The search space for optimal training configurations in \system is extensive, encompassing asymmetric combinations of PP and DP, model partitioning, 
interleaved pipelining, and microbatch sizes. 
Despite this complexity, \systemns's planner completes within 3 minutes for all 
workloads. This is a negligible overhead to pay for optimizing LLM training, which typically requires hundreds of thousands of GPU hours \cite{touvron2023llama}.
Fast planning is achieved through a two-phase optimization 
approach that employs approximation algorithms and heuristics to effectively prune the search space. 
Figure \ref{fig:optimizer_time_breakdown} presents a breakdown of optimization time for the largest models in each cluster.
While profiling dominates planning time, it scales sublinearly with the number of GPUs, since we make optimizations to parallelize 
profiling and avoid redundant measurements on duplicate VMs and GPUs.

\section{Related Work}
\noindent\textbf{Heterogeneous Training.~} Several systems have been proposed to optimize training on heterogeneous clusters with 
data parallelism (DP) \cite{moreno2020training,kim2022scale,whale,cannikin},
pipeline parallelism (PP) \cite{park2020hetpipe, pipepar, ding2021hetseq}, tensor parallelism (TP) \cite{hap}, and a combination of all three \cite{hexiscale,metis,fasop}. 
Existing strategies perform suboptimally when GPU memory and networking are limited. DP-only systems \cite{cephalo, zhang2025poplar, moreno2020training,kim2022scale} 
underutilize faster links as they are bottlenecked by slower links during collective communication. PP with DP reduces communication overhead 
but cannot efficiently utilize memory-efficient ZeRO-3 DP without significant communication costs.
TP \cite{hap, metis, hexiscale} is generally ineffective in heterogeneous environments due to high communication overhead (Section \ref{subsec:tp_het}). 
Systems combining approaches inherit limitations of each individual approach.
Systems like MiCS \cite{mics, holmes, zero++} optimize for heterogeneous networks, but lack support for clusters with compute and memory heterogeneity.
\system integrates PP and DP with optimizations enabling both communication and memory-efficient training while balancing resource heterogeneity,
delivering superior performance on heterogeneous clusters.

\noindent\textbf{Pipelining Optimizations.~} \cite{narayanan2021memory} introduces interleaved pipelining, using 1F1B scheduling that
interleaves forward and backward computations from different ministages.
When combined with ZeRO \cite{zero3}, this approach either incurs additional communication for parameter fetching,
or requires storing all ministage parameters in memory. \cite{zero_bubble} presents an optimized schedule
that eliminates pipeline bubbles, but requires keeping all ministage parameters in memory,
and without interleaving, optimizer updates cannot overlap with computation, incurring additional overhead.
\system uses interleaved pipelining with a GPipe-style \cite{huang2019gpipe} schedule, integrating with ZeRO-2 for communication efficiency while achieving
superior memory efficiency through parameter and activation offloading, and interleaved optimizer updates.

\noindent\textbf{Memory Optimizations.~}
Various works propose memory optimization strategies including activation checkpointing, activation offloading, and parameter offloading
\cite{cephalo,zero_offload,capuchin,checkmate,zero_infinity}. Mist \cite{zhu2025mist} jointly optimizes these techniques with parallelism
strategy, but does not address heterogeneous clusters. \system uniquely optimizes both checkpointing and offloading while targeting
heterogeneous training environments.

\vspace{-10pt}
\section{Discussion and Future Work}
\noindent\textbf{Dynamic Parallelism.~} Hardware performance variability and node failures present challenges, especially in cloud 
environments \cite{luo2022srifty, cloud_performance_variability,cloud_performance_variability_2}. 
Although \system determines parallelism configuration statically, its low-overhead planner enables periodic reconfiguration 
with updated performance models to accommodate hardware changes. \system could also incorporate existing approaches 
for efficient parallelism adaptation \cite{tenplex,hotspa,miao2023sdpipe} and node failure recovery \cite{parcae,recycle, varuna}.

\noindent\textbf{Tensor Parallelism.~} Although \system does not support TP due to its complexity and limited suitability for 
heterogeneous clusters, TP remains valuable for specific scenarios with very large models, extended context lengths, 
or small batch sizes. Future work could extend \system to incorporate DP + TP (including sequence parallelism \cite{sequence_parallelism}) combinations  
within GPU groups.

\noindent\textbf{Generalizability.~} Like previous studies \cite{cephalo,kim2022scale,hap,hexiscale,metis}, \system uses NVIDIA GPUs for 
evaluation due to their widespread availability and usage in training \cite{marketsandmarkets2023nvidia}. 
Nevertheless, \systemns's implementation is adaptable to any accelerator with PyTorch support, including AMD GPUs \cite{rocm_pytorch_install} 
and TPUs \cite{tpu,tpu_pytorch}. Furthermore, its design principles are hardware-agnostic and applicable 
to other heterogeneous environments.

\vspace{-10pt}
\section{Conclusion}
In this paper, we present \systemns, a system that allows users to train large language models efficiently on
clusters with significant compute, memory, and networking heterogeneity. 
\system achieves this by optimizing the combination of PP with DP such that it is both memory efficient and incurs low communication overhead
while balancing heterogeneous hardware utilization.
Moreover, it uses a planner to efficiently find an optimized training strategy from the vast search space of possible training configurations for 
heterogeneous clusters. Our experiments demonstrate that
\system can achieve up to 3$\times$ speedup in training throughput compared to existing state-of-the-art systems on three diverse
and representative heterogeneous training scenarios.

\bibliographystyle{ACM-Reference-Format}
\bibliography{main_arxiv}

%%
%% If your work has an appendix, this is the place to put it.
% \input{supplementary}

% \section{Research Methods}

% \subsection{Part One}

% Lorem ipsum dolor sit amet, consectetur adipiscing elit. Morbi
% malesuada, quam in pulvinar varius, metus nunc fermentum urna, id
% sollicitudin purus odio sit amet enim. Aliquam ullamcorper eu ipsum
% vel mollis. Curabitur quis dictum nisl. Phasellus vel semper risus, et
% lacinia dolor. Integer ultricies commodo sem nec semper.

\end{document}

%% file: plots/tp_vs_fsdp.tex
\begin{figure}
  \centering
  \begin{tikzpicture}
  \begin{groupplot}[
    group style={group size=1 by 1, vertical sep=25mm},
    ybar=0pt,
    height=4cm,
    legend style={at={(1.05,0.5)}, anchor=west, legend columns=1, font=\scriptsize},
    ylabel={HFU (\%)},
    nodes near coords,
    nodes near coords align={vertical},
    every node near coord/.append style={font=\tiny}, 
    nodes near coords={\pgfmathprintnumber[fixed,precision=0]{\pgfplotspointmeta}},
    ymajorgrids=true,
    grid style=dashed,
    ymin=0
  ]
  
  \nextgroupplot[
    width=6.7cm,
    height=3.8cm,
    symbolic x coords={T4,A10G,V100,A100},
    xtick=data,
    ymax=55,
    bar width=12pt,
    x tick label style={font=\scriptsize},
    y tick label style={font=\scriptsize},
    x tick style={draw=none},
    enlarge x limits=0.2
  ]
  \addplot+[style={fill=blue!70, pattern=north east lines, pattern color=blue!90}] 
      coordinates {(T4,13.14) (A10G,3.24) (V100,30.975) (A100,39.145)};
  \addplot+[style={fill=red!70, pattern=crosshatch, pattern color=red!90}] 
      coordinates {(T4,31.68) (A10G,32.28) (V100,44.97) (A100,44.46)};
  \legend{TP, DP (ZeRO-3)}
  \end{groupplot}
  \end{tikzpicture}
  \caption{Hardware Flops Utilization (HFU \cite{palm_hfu}) of tensor parallelism (TP) vs data parallelism (DP) with ZeRO-3 on 8 GPU AWS VMs.}
  \label{fig:tp_vs_fsdp}
  \vspace{-16pt}
\end{figure}

%% file: plots/network_bandwidth.tex
% Define reusable colors
\definecolor{myblue}{RGB}{34,94,168}
\definecolor{mybrown}{RGB}{204,76,2}

\begin{figure}[h]
  \centering
  \begin{subfigure}[b]{0.87\columnwidth}
    \centering
    \begin{tikzpicture}
      \begin{axis}[
        width=0.9\columnwidth,
        height=5.2cm,
        width=5.2cm,
        enlargelimits=false,
        colorbar,
        colorbar style={
          title=GB/s,
          ylabel shift=-20pt,
          title style={font=\tiny},
          tick label style={font=\tiny},
        },
        colormap={custom}{color(0)=(red!20!white); color(1)=(red!100!black)},
        xtick=data,
        ytick=data,
        xticklabels={\textcolor{myblue}{T4}, \textcolor{myblue}{A10G}, \textcolor{myblue}{V100}, \textcolor{myblue}{A100}, \textcolor{mybrown}{T4}, \textcolor{mybrown}{A10G}, \textcolor{mybrown}{V100}, \textcolor{mybrown}{A100}},
        yticklabels={\textcolor{myblue}{T4}, \textcolor{myblue}{A10G}, \textcolor{myblue}{V100}, \textcolor{myblue}{A100}, \textcolor{mybrown}{T4}, \textcolor{mybrown}{A10G}, \textcolor{mybrown}{V100}, \textcolor{mybrown}{A100}},
        xticklabel style={rotate=45, anchor=east, font=\tiny, yshift=-2pt},
        yticklabel style={font=\tiny},
        x tick style={draw=none},
        y tick style={draw=none},
        title style={font=\small},
        point meta min=0,
        point meta max=15,
        nodes near coords style={font=\tiny, text opacity=1},
        nodes near coords={\pgfmathprintnumber[fixed,precision=1]{\pgfplotspointmeta}},
        every node near coord/.append style={font=\tiny},
        nodes near coords style={text opacity=1, font=\tiny, anchor=center},
      ]
      \addplot[
  matrix plot,
  mesh/cols=8,
  point meta=explicit,
  shader=flat,              % use flat shading so draw works
  draw=black,               % border color
  line width=0.05pt,         % border thickness
  draw opacity=.8            % ensure the border is fully opaque
] table [meta=C] {
        x y C
        0 0 11.79
        1 0 12.06
        2 0 3.08
        3 0 11.76
        4 0 2.69
        5 0 2.69
        6 0 2.69
        7 0 2.69
        0 1 12.06
        1 1 12.16
        2 1 3.08
        3 1 12.26
        4 1 2.69
        5 1 2.69
        6 1 2.76
        7 1 2.69
        0 2 3.08
        1 2 3.08
        2 2 3.08
        3 2 3.08
        4 2 2.76
        5 2 2.69
        6 2 2.69
        7 2 2.69
        0 3 11.76
        1 3 12.26
        2 3 3.08
        3 3 12.26
        4 3 2.69
        5 3 2.69
        6 3 2.69
        7 3 2.69
        0 4 2.69
        1 4 2.69
        2 4 2.69
        3 4 2.69
        4 4 11.79
        5 4 12.06
        6 4 3.08
        7 4 11.76
        0 5 2.69
        1 5 2.69
        2 5 2.76
        3 5 2.69
        4 5 12.06
        5 5 12.16
        6 5 3.08
        7 5 12.26
        0 6 2.76
        1 6 2.69
        2 6 2.69
        3 6 2.69
        4 6 3.08
        5 6 3.08
        6 6 3.08
        7 6 3.08
        0 7 2.69
        1 7 2.69
        2 7 2.69
        3 7 2.69
        4 7 11.76
        5 7 12.26
        6 7 3.08
        7 7 12.26
      };
      \end{axis}
    \end{tikzpicture}
    \caption{Heatmap of inter-node bandwidth between VMs of different GPUs across two regions, \textcolor{myblue}{us-east-1} and \textcolor{mybrown}{us-east-2}.}
    \label{fig:internode_bandwidth}
  \end{subfigure}
  
  \begin{subfigure}[b]{0.9\columnwidth}
    \centering
    \begin{tikzpicture}
      \begin{axis}[
        width=0.9\columnwidth,
        height=3.5cm,
        ybar,
        ymax=265,
        ymin=0,
        enlarge x limits=0.15,
        ylabel={Bandwidth (GB/s)},
        ylabel style={font=\scriptsize},
        symbolic x coords={T4,A10G,V100,A100},
        xtick=data,
        yticklabel style={font=\scriptsize},
        xticklabel style={font=\tiny},
        x tick style={draw=none},
        ytick={},
        title style={font=\small},
        nodes near coords,
        nodes near coords style={font=\tiny},
      ]
      \addplot+[fill=myblue, bar width=15pt] coordinates {
        (T4,6.1)
        (A10G,3.0)
        (V100,23.9)
        (A100,222.2)
      };
      \end{axis}
    \end{tikzpicture}
    \caption{Intra-node bandwidth within VMs.}
    \label{fig:intranode_bandwidth}
  \end{subfigure}
  \caption{Unidirectional bandwidth between GPUs on AWS.}
  \label{fig:network_bandwidth}
  \vspace{-2pt}
\end{figure}

%% file: plots/zero2_vs_zero3.tex
\begin{figure}[h]
  \centering
  \begin{tikzpicture}
  \begin{groupplot}[
    group style={group size=2 by 1, horizontal sep=10mm},
    ybar,
    ymin=0,
    enlarge x limits=0.35,
    width=0.5\columnwidth,
    xticklabel style={font=\tiny},
    y tick label style={font=\tiny},
    x tick style={draw=none},
    ylabel style={font=\scriptsize},
    xlabel style={font=\scriptsize},
    tick label style={font=\scriptsize},
    xticklabel style={rotate=20, anchor=east, font=\tiny, yshift=-2.4pt, xshift=15},
    legend style={at={(1.05,0.5)}, anchor=west, legend columns=1, font=\tiny},
    height=4cm,
    width=0.47\columnwidth,
    % nodes near coords,
    % nodes near coords align={vertical},
    % every node near coord/.append style={font=\tiny, anchor=west}
  ]
  
  % Plot 1: TFlops
  \nextgroupplot[
    ylabel={TFlops},
    symbolic x coords={Llama 1B, Llama 3B, Llama 7B},
    xtick=data
  ]
  \addplot+[pattern color={rgb:red, 152; green, 78; blue, 163}, pattern=north east lines, bar width=5pt] coordinates {
    (Llama 1B, 329.2)
    (Llama 3B, 249.58)
    (Llama 7B, 0)
  };
  \addplot+[pattern color={rgb:red, 255; green, 127; blue, 0}, pattern=crosshatch, bar width=5pt] coordinates {
    (Llama 1B, 61.99)
    (Llama 3B, 119.07)
    (Llama 7B, 163.31)
  };
  
  % Add OOM label for Llama 7B with PP+ZeRO-2
  \node[text=red, align=center, font=\tiny, rotate=90,yshift=5pt] at (axis cs:Llama 7B, 50) {OOM};
  
  % Plot 2: Memory Utilization
  \nextgroupplot[
    ylabel={Memory Util. (\%)},
    symbolic x coords={Llama 1B, Llama 3B, Llama 7B},
    xtick=data
  ]
  \addplot+[pattern color={rgb:red, 152; green, 78; blue, 163}, pattern=north east lines, bar width=5pt] coordinates {
    (Llama 1B, 36.98)
    (Llama 3B, 68.02)
    (Llama 7B, 0)
  };
  \addplot+[pattern color={rgb:red, 255; green, 127; blue, 0}, pattern=crosshatch, bar width=5pt] coordinates {
    (Llama 1B, 14.32)
    (Llama 3B, 31.33)
    (Llama 7B, 48.98)
  };
  
  % Add OOM label for Llama 7B with PP+ZeRO-2
  \node[text=red, align=center, font=\tiny, rotate=90,yshift=5pt] at (axis cs:Llama 7B, 10) {OOM};
  
  \legend{PP+ZeRO-2, PP+ZeRO-3}
  \end{groupplot}
  \end{tikzpicture}
  \caption{Comparison of pipeline parallelism (PP) with ZeRO-2 vs ZeRO-3 across different Llama\cite{touvron2023llama, zhang2024tinyllama} 
  model sizes on 8 V100s + 8 T4s. OOM indicates Out-of-Memory.}
  \label{fig:zero2_vs_zero3}
\end{figure} 

%% file: tables/system_comparison.tex
\begin{table}[h]
  \centering
  \renewcommand{\arraystretch}{1.1}
  \footnotesize % Reduces the font size
  \resizebox{\linewidth}{!}{%
  \begin{tabular}{l c c c c c}
      \toprule
      System & Communication & Memory  & Computation & PP & Asymmetric \\
             & Efficient DP & Efficient DP & Balancing DP & +DP & PP \\
      \midrule
      TorchTitan-ZeRO2 \cite{torchtitan} & \textbf{\checkmark} & \textbf{\Large\xmark} & \textbf{\Large\xmark} & (\textbf{\checkmark}) & \textbf{\Large\xmark} \\
      TorchTitan-ZeRO3 \cite{torchtitan} & \textbf{\Large\xmark} & \textbf{\checkmark} & \textbf{\Large\xmark} & (\textbf{\checkmark}) & \textbf{\Large\xmark} \\ 
      HexiScale \cite{hexiscale} & \textbf{\checkmark} & \textbf{\Large\xmark} & (\textbf{\checkmark}) & (\textbf{\checkmark}) & \textbf{\checkmark} \\
      Cephalo \cite{cephalo} & \textbf{\checkmark} & \textbf{\checkmark} & \textbf{\checkmark} & \textbf{\Large\xmark} & \textbf{\Large\xmark} \\\hline
      \textbf{\system (Our System)} & \textbf{\checkmark} & \textbf{\checkmark} & \textbf{\checkmark} & \textbf{\checkmark} & \textbf{\checkmark} \\ 
      \bottomrule
  \end{tabular}
  }
  \textbf{\checkmark} Supported, (\textbf{\checkmark}) Supported with manual tuning, \textbf{\Large\xmark} Not supported
  \caption{Comparison of heterogeneous training systems.}
  \label{tab:system_comparison_table}
  \vspace{-5pt}
\end{table} 

% \toprule
% System & Communication & Memory  & Computation & Asymmetric \\
%         & Efficient DP & Efficient DP & Balancing DP & Pipeline Parallelism \\
% \midrule
% TorchTitan-ZeRO2 \cite{torchtitan} & \checkmark & \xmark & \xmark & \xmark \\
% TorchTitan-ZeRO3 \cite{torchtitan} & \xmark & \checkmark & \xmark & \xmark \\ 
% HexiScale \cite{hexiscale} & \checkmark & \xmark & \checkmark & \checkmark \\
% Cephalo \cite{cephalo} & \checkmark & \checkmark & \checkmark & \xmark \\\hline
% \textbf{\system (Our System)} & \checkmark & \checkmark & \checkmark & \checkmark \\ 
% \bottomrule

%% file: tables/gpu_specs.tex
\begin{table}[ht]
  \centering
  \caption{Datacenter-class GPU Specifications}
  \vspace{0.2em}
  \renewcommand{\arraystretch}{0.95}
  \setlength{\extrarowheight}{0.5pt}
  \resizebox{0.75\linewidth}{!}{%
  \begin{tabular}{|c|l|l|c|c|}
    \hline
    \textbf{Performance} & \textbf{GPU}      & \textbf{Memory} & \textbf{TFlops (FP16)}  \\ \hline
    \multirow{2}{*}{High} & H100-NVL & 94 GB         & 989            \\
    & A100      & 40/80 GB         & 312         \\ \hline
    \multirow{2}{*}{Middle}
    & V100      & 16 GB         & 125         \\
    & A10G      & 24 GB         & 125         \\ \hline
   Low &  T4        & 16 GB         & 65           \\ \hline
  \end{tabular}
  }
  \label{tab:gpu_specs}
  \vspace{-13pt}
\end{table} 

% \textbf{Performance} & \textbf{GPU}      & \textbf{Memory} & \textbf{TFlops (FP16)} & \textbf{Mem. Bandwidth} \\ \hline
% \multirow{2}{*}{High} & H100-NVL & 94 GB         & 989            & 3900 GB/s               \\
% & A100      & 40/80 GB         & 312         & 1555 GB/s                \\ \hline
% \multirow{3}{*}{Middle}
% & V100      & 16 GB         & 125         & 900 GB/s                 \\
% & A10G      & 24 GB         & 125         & 600 GB/s                 \\
% & L4        & 24 GB         & 120           & 300 GB/s                 \\ \hline
% Low &  T4        & 16 GB         & 65           & 320 GB/s                 \\ \hline

%% file: tables/cluster_gpu_configs.tex
\begin{table}[ht]
  \centering
  \caption{Cluster GPU Configurations}
  \vspace{0.2em}
  \renewcommand{\arraystretch}{1.05}
  \setlength{\extrarowheight}{0.5pt}
  \resizebox{0.95\linewidth}{!}{%
  \begin{tabular}{|c|c|c|c|c|c|c|}
  \hline
  \textbf{Cluster} & \textbf{Cloud} & \textbf{\# VMs} & \textbf{GPUs / VM} & \textbf{Total} & \textbf{Total} & \textbf{Total}\\ 
& & & & \textbf{Regions} & \textbf{GPUs} & \textbf{TFlops (FP16)} \\ \hline
  \multirow{2}{*}{A} & \multirow{2}{*}{Azure} & 2 & 2$\times$H100 & \multirow{2}{*}{1} & \multirow{2}{*}{20} & \multirow{2}{*}{8332} \\ 
    &  & 2 & 8$\times$A100 (80GB) &  &  &  \\ \hline
  \multirow{4}{*}{B} 
  & \multirow{4}{*}{AWS} & 1 & 8$\times$A100 (40GB)
  & \multirow{4}{*}{1} &
   \multirow{4}{*}{64} & \multirow{4}{*}{8112} \\ 
    &  & 2 & 8$\times$A10G &  &  &  \\ 
    &  & 2 & 8$\times$V100 &  &  &  \\ 
    &  & 3 & 8$\times$T4 &  &  &  \\ \hline
  \multirow{3}{*}{C} & \multirow{3}{*}{AWS} & 2 & 8$\times$A10G & \multirow{3}{*}{2} & \multirow{3}{*}{128} & \multirow{3}{*}{8240} \\ 
    &  & 2 & 8$\times$V100 &  &  &  \\ 
    &  & 12 & 8$\times$T4 &  &  &  \\ \hline
  \end{tabular}
  }
  \label{tab:cluster_gpu_configs}
  \vspace{-10pt}
\end{table} 

%% file: tables/training_throughput_results.tex
\begin{table*}[ht]
  \centering
  \scriptsize
  \caption{Throughput (TFlops) and GPU utilization (HFU) of \system compared to other systems across different model sizes and clusters (higher is better). \textit{OOM} denotes Out-of-Memory.}
  \label{table:training_throughput_results}
  \renewcommand{\arraystretch}{0.4}
  \resizebox{0.88\textwidth}{!}{
  \begin{tabular}{@{}llrrrrrrrr@{}}
  \toprule
  \textbf{Cluster} & \textbf{Model} & \multicolumn{2}{c}{\textbf{\systemns}} 
  & \multicolumn{2}{c}{\textbf{TorchTitan-Het}} & \multicolumn{2}{c}{\textbf{HexiScale}} 
  & \multicolumn{2}{c}{\textbf{Cephalo}} \\
\cmidrule(lr){3-4} \cmidrule(lr){5-6} \cmidrule(lr){7-8} \cmidrule(lr){9-10}
& & \textbf{TFlops} & \textbf{HFU} & \textbf{TFlops} & \textbf{HFU} & \textbf{TFlops} & \textbf{HFU} & \textbf{TFlops} & \textbf{HFU} \\
\midrule
\multirow{4}{*}{A} 
& Llama 7B  & \textbf{4370.56} & \textbf{52.46\%} & 4223.80 & 50.69\% & 3193.46 & 38.33\% & 1714.52 & 20.58\% \\
& Llama 13B & \textbf{4917.87} & \textbf{59.02\%} & 3837.49 & 46.06\% & 3270.32 & 39.25\% & 1656.29 & 19.88\% \\
& Llama 33B & \textbf{5281.64} & \textbf{63.39\%} &  944.47 & 11.34\% & 3064.22 & 36.78\% & 1943.89 & 23.33\% \\
& Llama 65B & \textbf{5239.13} & \textbf{62.88\%} & \textit{OOM} & \textit{OOM} & 2048.63 & 24.59\% & 1937.64 & 23.26\% \\
  \midrule
  \multirow{3}{*}{B} & Llama 7B  & \textbf{3412.88} & \textbf{43.49\%} & 2033.53 & 25.91\% & 1194.89 & 15.23\% & 2274.50 & 28.98\% \\
                    & Llama 13B & \textbf{2965.64} & \textbf{37.79\%} & 1956.09 & 24.93\% & 1152.73 & 14.69\% & 1992.24 & 25.39\% \\
                    & Llama 33B & \textbf{2658.29} & \textbf{33.87\%} & \textit{OOM} & \textit{OOM} &  657.16 &  8.37\% & 1373.31 & 17.50\% \\
  \midrule
  \multirow{3}{*}{C} & Llama 7B  & \textbf{3936.94} & \textbf{39.24\%} & 2441.70 & 24.34\% & 2624.63 & 26.16\% & 1213.39 & 12.10\% \\
                    & Llama 13B & \textbf{3357.97} & \textbf{33.47\%} & 2061.55 & 20.55\% & 1952.31 &	19.46\% & 1222.96 & 12.19\% \\
                    & Llama 33B & \textbf{1548.60} & \textbf{15.44\%} & \textit{OOM} & \textit{OOM} &  \textit{OOM} &  \textit{OOM} &  775.42 &  7.73\% \\
  \bottomrule
  \end{tabular}
  }
  \vspace{-5pt}
\end{table*} 

%% file: plots/cluster_scaling.tex
% Define ColorBrewer Set1 colors
\definecolor{cbBlue}{RGB}{55,126,184}   % PFlops
\definecolor{cbRed}{RGB}{228,26,28}     % HFU

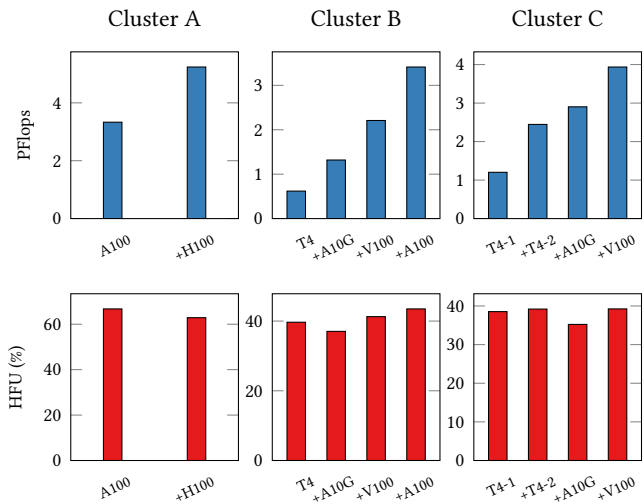
\begin{figure}[h]
  \centering
  \begin{tikzpicture}
  \begin{groupplot}[
    group style={group size=3 by 2, horizontal sep=4.5mm, vertical sep=10mm},
    ybar,
    ymin=0,
    enlarge x limits=0.20,
    width=0.45\columnwidth,
    y tick label style={font=\scriptsize},
    x tick style={draw=none, font=\tiny},
    ylabel style={font=\scriptsize},
    xlabel style={font=\tiny},
    tick label style={font=\tiny},
    title style={font=\small},
    height=3.8cm,
    xticklabel style={rotate=25, anchor=north east, font=\tiny, xshift=8pt},
  ]
  
  % PFlops Cluster A
  \nextgroupplot[
    ylabel={PFlops},
    title={Cluster A},
    symbolic x coords={A100,+H100},
    enlarge x limits=0.50,
    xtick=data,
    bar width=7pt,
  ]
  \addplot+[fill=cbBlue, draw=black] coordinates {
    (A100, 3.33152)
    (+H100, 5.23913)
  };
  
  % PFlops Cluster B
  \nextgroupplot[
    title={Cluster B},
    symbolic x coords={T4,+A10G,+V100,+A100},
    xtick=data,
    bar width=7pt,
  ]
  \addplot+[fill=cbBlue, draw=black] coordinates {
    (T4, 0.6189)
    (+A10G, 1.31888)
    (+V100, 2.20872)
    (+A100, 3.41288)
  };
  
  % PFlops Cluster C
  \nextgroupplot[
    title={Cluster C},
    symbolic x coords={T4-1,+T4-2,+A10G,+V100},
    xtick=data,
    bar width=7pt,
  ]
  \addplot+[fill=cbBlue, draw=black] coordinates {
    (T4-1, 1.20251)
    (+T4-2, 2.44595)
    (+A10G, 2.90285)
    (+V100, 3.93694)
  };
  
  % HFU Cluster A
  \nextgroupplot[
    ylabel={HFU (\%)},
    symbolic x coords={A100,+H100},
    xtick=data,
    enlarge x limits=0.50,
    bar width=7pt,
  ]
  \addplot+[fill=cbRed, draw=black] coordinates {
    (A100, 66.74)
    (+H100, 62.88)
  };
  
  % HFU Cluster B
  \nextgroupplot[
    symbolic x coords={T4,+A10G,+V100,+A100},
    xtick=data,
    bar width=7pt,
  ]
  \addplot+[fill=cbRed, draw=black] coordinates {
    (T4, 39.67)
    (+A10G, 37.05)
    (+V100, 41.27)
    (+A100, 43.49)
  };
  
  % HFU Cluster C
  \nextgroupplot[
    symbolic x coords={T4-1,+T4-2,+A10G,+V100},
    xtick=data,
    bar width=7pt,
  ]
  \addplot+[fill=cbRed, draw=black] coordinates {
    (T4-1, 38.54)
    (+T4-2, 39.20)
    (+A10G, 35.23)
    (+V100, 39.24)
  };
  
  \end{groupplot}
  \end{tikzpicture}
  \caption{PFlops (1k TFlops) and HFU scaling as heterogeneous GPUs are added to the clusters.}
  \label{fig:cluster_scaling}
  \vspace{-12pt}
\end{figure}

%% file: plots/homogeneous_comparison.tex
\begin{figure*}
  \centering
  \begin{tikzpicture}
  \begin{groupplot}[
    group style={group size=3 by 1, horizontal sep=14mm},
    ybar=0pt,
    height=3.2cm,
    legend style={at={(0.5,-0.3)}, anchor=north, legend columns=-1, font=\scriptsize},
    ylabel={HFU (\%)},
    nodes near coords,
    nodes near coords align={vertical},
    title style={font=\scriptsize},
    y tick label style={font=\scriptsize},
    x tick style={draw=none},
    ylabel style={font=\scriptsize},
    every node near coord/.append style={font=\tiny, yshift=-0.5mm}, 
    ymajorgrids=true,
    grid style=dashed,
    ymin=0
  ]
  
  % Plot 1: Cluster A (H100, A100)
  \nextgroupplot[
    title={Llama 13B, Cluster A},
    width=5cm,
    symbolic x coords={A100,H100},
    xtick=data,
    bar width=9pt,
    ymax=80,
    x tick label style={font=\scriptsize},
    enlarge x limits=0.45
  ]
  \addplot+[pattern color={rgb,255:red, 0; green, 107; blue, 164}, pattern=crosshatch dots] 
      coordinates {(A100,64) (H100,52)};
  \addplot+[pattern color={rgb,255:red, 255; green, 128; blue, 14}, pattern=north east lines] 
      coordinates {(A100,67) (H100,57)};
  
  % Plot 2: Cluster B (T4, A10G, V100, A100)
  \nextgroupplot[
    title={Llama 7B, Cluster B},
    width=6.0cm,
    bar width=9pt,
    symbolic x coords={T4,A10G,V100,A100},
    xtick=data,
    ymax=75,
    x tick label style={font=\scriptsize},
    enlarge x limits=0.3
  ]
  \addplot+[pattern color={rgb,255:red, 0; green, 107; blue, 164}, pattern=crosshatch dots] 
      coordinates {(T4,30) (A10G,33) (V100,51) (A100,56)};
  \addplot+[pattern color={rgb,255:red, 255; green, 128; blue, 14}, pattern=north east lines] 
      coordinates {(T4,40) (A10G,39) (V100,56) (A100,59)};
  \legend{Heterogeneous, Homogeneous}
  
  % Plot 3: Cluster C (T4, A10G, V100)
  \nextgroupplot[
    title={Llama 7B, Cluster C},
    width=5.5cm,
    bar width=9pt,
    symbolic x coords={T4,A10G,V100},
    xtick=data,
    ymax=70,
    x tick label style={font=\scriptsize},
    enlarge x limits=0.3
  ]
  \addplot+[pattern color={rgb,255:red, 0; green, 107; blue, 164}, pattern=crosshatch dots] 
      coordinates {(T4,35) (A10G,35) (V100,58)};
  \addplot+[pattern color={rgb,255:red, 255; green, 128; blue, 14}, pattern=north east lines] 
      coordinates {(T4,39) (A10G,39) (V100,56)};
  \end{groupplot}
  \end{tikzpicture}
  \caption{GPU Utilization on heterogeneous clusters vs homogeneous subgroups of GPUs within the clusters.}
  \label{fig:homog_hfu}
  \vspace{-8pt}
\end{figure*}
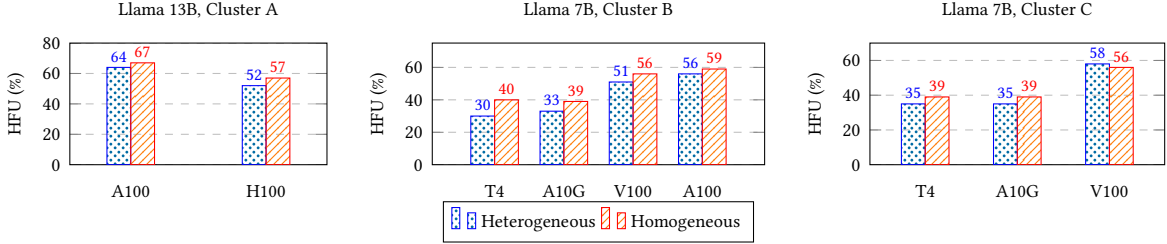 

% \addplot+[style={fill=blue!70}] 
% coordinates {(A100,63.54) (H100,52.28)};
% \addplot+[style={fill=red!70}] 
% coordinates {(A100,66.74) (H100,56.91)};
% \legend{Heterogeneous, Homogeneous}

% % Plot 2: Cluster B (T4, A10G, V100, A100)
% \nextgroupplot[
% title={Llama 7B, Cluster B},
% width=7cm,
% bar width=12pt,
% symbolic x coords={T4,A10G,V100,A100},
% xtick=data,
% ymax=70,
% x tick label style={font=\small},
% enlarge x limits=0.3
% ]
% \addplot+[style={fill=blue!70}] 
% coordinates {(T4,29.96) (A10G,32.96) (V100,50.53) (A100,56.25)};
% \addplot+[style={fill=red!70}] 
% coordinates {(T4,39.67) (A10G,38.69) (V100,56.01) (A100,58.61)};

%% file: plots/interleaved_pipelining.tex
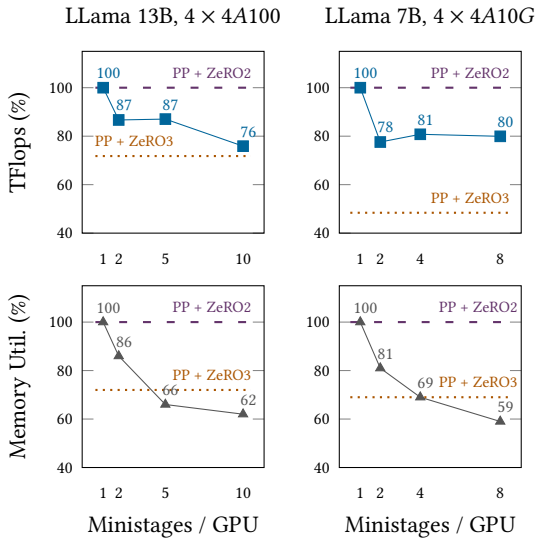
\begin{figure}[h]
  \centering
  \begin{tikzpicture}
  \begin{groupplot}[
    group style={group size=2 by 2, horizontal sep=10mm, vertical sep=7mm},
    ymin=40,
    ymax=115,
    enlarge x limits=0.15,
    width=0.45\columnwidth,
    xticklabel style={rotate=0, anchor=north, font=\tiny, yshift=-4pt},
    y tick label style={font=\tiny},
    x tick style={draw=none},
    ylabel style={font=\small},
    xlabel style={font=\small},
    ymajorgrids=false,
    tick label style={font=\small},
    height=4cm,
    width=4cm,
    title style={font=\small},
    nodes near coords={\pgfmathprintnumber[fixed,precision=0]{\pgfplotspointmeta}},
    every node near coord/.append style={font=\tiny, xshift=2pt}
  ]

  % Plot 1: TFlops for LLama 13B (top left)
  \nextgroupplot[
    ylabel={TFlops (\%)},
    title={LLama 13B, $4\times4 A100$},
    xtick={1,2,5,10},
  ]
  \addplot[mark=square*, color={rgb:red, 0; green, 107; blue, 164}] coordinates {
    (1, 100)
    (2, 86.69)
    (5, 87.05)
    (10, 75.88)
  };
  \draw[color={rgb:red, 255; green, 127; blue, 0}, dotted, thick] (axis cs:0.5,71.8) -- (axis cs:10.5,71.8) node[pos=0.25, above, font=\tiny] {PP + ZeRO3};
  \draw[color={rgb:red, 152; green, 78; blue, 163}, loosely dashed, thick] (axis cs:0.5,100) -- (axis cs:10.5,100) node[pos=0.75, above, font=\tiny] {PP + ZeRO2};

  % Plot 2: TFlops for LLama 7B (top right)
  \nextgroupplot[
    title={LLama 7B, $4\times4 A10G$},
    xtick={1,2,4,8},
  ]
  \addplot[mark=square*, color={rgb:red, 0; green, 107; blue, 164}] coordinates {
    (1, 100)
    (2, 77.59)
    (4, 80.78)
    (8, 79.94)
  };
  \draw[color={rgb:red, 255; green, 127; blue, 0}, dotted, thick] (axis cs:0.5,48.4) -- (axis cs:8.5,48.4) node[pos=0.75, above, font=\tiny] {PP + ZeRO3};
  \draw[color={rgb:red, 152; green, 78; blue, 163}, loosely dashed, thick] (axis cs:0.5,100) -- (axis cs:8.5,100) node[pos=0.75, above, font=\tiny] {PP + ZeRO2};
  
  % Plot 3: Memory Utilization for LLama 13B (bottom left)
  \nextgroupplot[
    ylabel={Memory Util. (\%)},
    xlabel={Ministages / GPU},
    xtick={1,2,5,10},
  ]
  \addplot[mark=triangle*, color={rgb:red, 89; green, 89; blue, 89}] coordinates {
    (1, 100)
    (2, 86)
    (5, 66)
    (10, 62)
  };
  \draw[color={rgb:red, 255; green, 127; blue, 0}, dotted, thick] (axis cs:0.5,72) -- (axis cs:10.5,72) node[pos=0.75, above, font=\tiny] {PP + ZeRO3};
  \draw[color={rgb:red, 152; green, 78; blue, 163}, loosely dashed, thick] (axis cs:0.5,100) -- (axis cs:10.5,100) node[pos=0.75, above, font=\tiny] {PP + ZeRO2};

  % Plot 4: Memory Utilization for LLama 7B (bottom right)
  \nextgroupplot[
    xlabel={Ministages / GPU},
    xtick={1,2,4,8},
  ]
  \addplot[mark=triangle*, color={rgb:red, 89; green, 89; blue, 89}] coordinates {
    (1, 100)
    (2, 81)
    (4, 69)
    (8, 59)
  };
  \draw[color={rgb:red, 255; green, 127; blue, 0}, dotted, thick] (axis cs:0.5,69) -- (axis cs:8.5,69) node[pos=0.80, above, font=\tiny] {PP + ZeRO3};
  \draw[color={rgb:red, 152; green, 78; blue, 163}, loosely dashed, thick] (axis cs:0.5,100) -- (axis cs:8.5,100) node[pos=0.80, above, font=\tiny] {PP + ZeRO2};

  \end{groupplot}
  \end{tikzpicture}
  \caption{TFlops and memory utilization for varying ministages per GPU. Values are normalized to 1 microstage.}
  \label{fig:interleaved_pipelining_scaling}
  \vspace{-15pt}
\end{figure} 
% \addplot[mark=*, blue] coordinates {
%     (1, 1.00)
%     (2, 0.8669413929)
%     (5, 0.8705250739)
%     (10, 0.758761148)
%   };

%   % Plot 2: Memory Utilization for LLama 13B (top right)
%   \nextgroupplot[
%     ylabel={Memory Util.},
%     xtick={1,2,5,10},
%     xlabel={Microstages / GPU},
%   ]
%   \addplot[mark=*, red] coordinates {
%     (1, 1.00)
%     (2, 0.86)
%     (5, 0.66)
%     (10, 0.62)
%   };
  
%   % Plot 3: TFlops for LLama 7B (bottom left)
%   \nextgroupplot[
%     ylabel={TFlops},
%     xlabel={Microstages / GPU},
%     title={LLama 7B, $4\times4 A10G$},
%     xtick={1,2,4,8},
%   ]
%   \addplot[mark=*, blue] coordinates {
%     (1, 1.00)
%     (2, 0.7759342916)
%     (4, 0.8077700205)
%     (8, 0.79937577)
%   };

%   % Plot 4: Memory Utilization for LLama 7B (bottom right)
%   \nextgroupplot[
%     ylabel={Memory Util.},
%     xlabel={Microstages / GPU},
%     xtick={1,2,4,8},
%   ]
%   \addplot[mark=*, red] coordinates {
%     (1, 1.00)
%     (2, 0.81)
%     (4, 0.69)
%     (8, 0.59)
%   };

%% file: plots/ablation_study.tex
\definecolor{cbBlue}{RGB}{55,126,184}   % TFlops
\definecolor{cbGrey}{RGB}{89,89,89} 
% {rgb,255:red, 89; green, 89; blue, 89}    % Memory Util.

\begin{figure}[h]
  \centering
  \begin{tikzpicture}
  \begin{groupplot}[
    group style={group size=2 by 1, horizontal sep=15mm},
    ybar,
    symbolic x coords={Baseline,Baseline+O,Baseline+O+I,Baseline+O+I+H},
    xtick=data,
    ymin=0,
    enlarge x limits=0.15,
    width=0.5\columnwidth,
    xticklabel style={rotate=20, anchor=east, font=\tiny, yshift=-4pt, xshift=5},
    y tick label style={font=\tiny},
    x tick style={draw=none},
    ylabel style={font=\small},
    xlabel style={font=\small},
    tick label style={font=\small},
    height=4cm
  ]
  
  % Plot 1: TFlops
  \nextgroupplot[
    ylabel={TFlops}
  ]
  \addplot+[fill=cbBlue, draw=black] coordinates {
    (Baseline, 0)
    (Baseline+O, 3746.11)
    (Baseline+O+I, 4086.89)
    (Baseline+O+I+H, 5239.13)
  };
  \node[text width=1cm, align=center, font=\tiny, text=red] at (axis cs:Baseline,700) {OOM};
  
  % Plot 2: Memory
  \nextgroupplot[
    ylabel={Memory Util. (\%)}
  ]
  \addplot+[fill=cbGrey, draw=black] coordinates {
    (Baseline, 0)
    (Baseline+O, 83.12)
    (Baseline+O+I, 57.89)
    (Baseline+O+I+H, 72.91)
  };
  \node[text width=1cm, align=center, font=\tiny, text=red] at (axis cs:Baseline,10) {OOM};
  
  \end{groupplot}
  \end{tikzpicture}
  \caption{TFlops and memory utilization of \system optimizations (Cluster A, LLama 65B). Baseline runs out of memory.}
  \label{fig:ablation_study}
  \vspace{-18pt}
\end{figure}
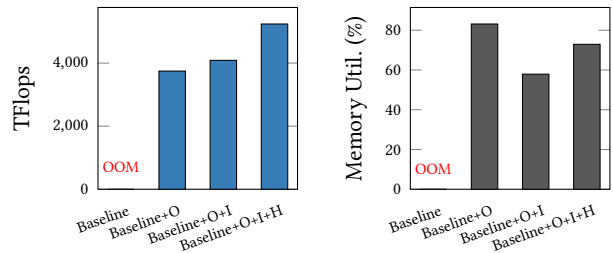 

%% file: plots/optimizer_time_breakdown.tex
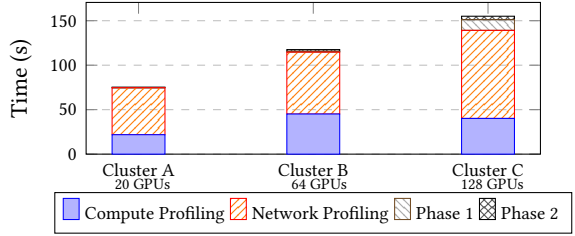
\begin{figure}
  \centering
  \begin{tikzpicture}
  \begin{axis}[
      ybar stacked,
      bar width=20pt,
      height=3.6cm,
      width=5cm,
      width=0.9\columnwidth,
      enlarge x limits=0.15,
      ylabel={Time (s)},
      symbolic x coords={Cluster A, Cluster B, Cluster C},
      xtick=data,
      legend style={
          at={(0.5,-0.25)},
          anchor=north,
          legend columns=-1,
          font=\scriptsize
      },
      ymin=0,
      y tick label style={font=\scriptsize},
      ylabel style={font=\small},
      x tick label style={font=\scriptsize, draw=none},
      ymajorgrids=true,
      grid style=dashed,
      nodes near coords align={below},
      every node near coord/.append style={font=\scriptsize}
  ]
  % Color-blind friendly colors
  \addplot+[pattern color={rgb,255:red, 0; green, 107; blue, 164}]     
  coordinates {(Cluster A,21.86) (Cluster B,45.12) (Cluster C,40.16)}; % Blue
  \addplot+[pattern color={rgb,255:red, 255; green, 128; blue, 14}, pattern=north east lines]    
  coordinates {(Cluster A,52.86) (Cluster B,69.62) (Cluster C,99.02)}; % Orange
  \addplot+[pattern color={rgb,255:red, 171; green, 171; blue, 171}, pattern=north west lines]   
  coordinates {(Cluster A,0.01) (Cluster B,0.89) (Cluster C,12.03)}; % Grey
  \addplot+[pattern color={rgb,255:red, 89; green, 89; blue, 89}, pattern=crosshatch]      
  coordinates {(Cluster A,0.6) (Cluster B,1.92) (Cluster C,3.84)}; % Dark Grey
  \legend{Compute Profiling, Network Profiling, Phase 1, Phase 2}
  \end{axis}
  
  % Add GPU count labels
  \node[font=\tiny] at (0.75, -0.4) {20 GPUs};
  \node[font=\tiny] at (3.1, -0.4) {64 GPUs};
  \node[font=\tiny] at (5.4, -0.4) {128 GPUs};
  \end{tikzpicture}
  \caption{Planner runtime breakdown per cluster.}
  \label{fig:optimizer_time_breakdown}
  \vspace{-10pt}
\end{figure}